\documentclass[review]{elsarticle}
\usepackage{booktabs}
\usepackage{graphicx}
\usepackage{amsmath,amssymb,amsfonts}
\usepackage{units}
\usepackage{epstopdf}
\usepackage{caption}
\usepackage{color}
\usepackage{hyperref}
\usepackage{pbox}
\usepackage{xspace}
\usepackage{amsmath}
\usepackage{upgreek}
\usepackage{footnote}
\usepackage{rotating}

\usepackage{etoolbox}
\makeatletter
    \patchcmd{\tnotemark}{\ding{73}}{\dag,}{}{\@latex@error{Failed to path \string\tnotemark\space for \string\ding{73}}}
    \patchcmd{\tnotetext}{\ding{73}}{\dag}{}{\@latex@error{Failed to path \string\tnotetext\space for \string\ding{73}}}
\makeatother

\hypersetup{
    colorlinks,
    citecolor=black,
    filecolor=black,
    linkcolor=black,
    urlcolor=black
}
\newcommand{\Nuc}[2]{\ensuremath{^{#2}\mbox{#1}}}
\DeclareGraphicsExtensions{.pdf}
\journal{Astroparticle Physics}

\begin{document}

\title{Design and Construction of the DEAP-3600 Dark Matter Detector}
\author[triumf]{P.-A.~Amaudruz}
\author[ral]{M.~Baldwin}
\author[laurentian]{M.~Batygov}
\author[alberta]{B.~Beltran}
\author[alberta]{C.\,E.~Bina}
\author[triumf]{D.~Bishop}
\author[queens]{J.~Bonatt}
\author[rhul]{G.~Boorman}
\author[queens,carleton]{M.\,G.~Boulay}
\author[queens]{B.~Broerman}
% ...

\address[alberta]{Department of Physics, University of Alberta, \\Edmonton, Alberta, T6G 2R3, Canada}
\address[cr]{Canadian Nuclear Laboratories Ltd., Chalk River Laboratories, \\Chalk River, K0J 1P0 Canada}
\address[carleton]{Department of Physics, Carleton University, \\Ottawa, Ontario, K1S 5B6, Canada}
\address[laurentian]{Department of Physics and Astronomy, Laurentian University, \\Sudbury, Ontario, P3E 2C6, Canada}
\address[unam]{Instituto de F\'isica Universidad Nacional Aut\'onoma de M\'exico,\\ Apartado Postal 20-364, M\'exico D. F. 01000}
\address[queens]{Department of Physics, Engineering Physics, and Astronomy, Queen's University, \\Kingston, Ontario, K7L 3N6, Canada}
\address[snolab]{SNOLAB, Lively, Ontario, P3Y 1M3, Canada}
\address[sussex]{Department of Physics and Astronomy, University of Sussex, \\Sussex House, Brighton, East Sussex BN1 9RH, United Kingdom}
\address[rhul]{Department of Physics, Royal Holloway, University of London, \\Egham Hill, Egham, Surrey TW20 0EX, United Kingdom}
\address[ral]{Rutherford Appleton Laboratories, Swindon SN2 1SZ, United Kingdom}
\address[tum]{Department of Physics, Technische Universit\"at M\"unchen, \\80333 Munich, Germany}
\address[triumf]{TRIUMF, Vancouver, British Columbia, V6T 2A3, Canada}

\begin{abstract}
The Dark matter Experiment using Argon Pulse-shape discrimination (DEAP) has been designed for a direct detection search for particle dark matter using a single-phase liquid argon target. The projected cross section sensitivity for DEAP-3600 to the spin-independent scattering of Weakly Interacting Massive Particles (WIMPs) on nucleons is $10^{-46}~\rm{cm}^{2}$ for a 100~GeV/$c^2$ WIMP mass with a fiducial exposure of 3~tonne-years. This paper describes the physical properties and construction of the DEAP-3600 detector. 
\end{abstract}
\maketitle
\newpage

\tableofcontents
\addtocontents{toc}{\protect\setcounter{tocdepth}{2}}

\section{Introduction}\label{sec:introduction}
The origin of dark matter in the universe is one of the most important questions in particle astrophysics. A well-motivated dark matter candidate is the Weakly Interacting Massive Particle (WIMP), which is predicted naturally in supersymmetric extensions of the Standard Model~\cite{jungman1996supersymmetric, bertone2005particle}. To date, WIMPs remain undetected in laboratory-based searches~\cite{aprile2017first, agnes2016results, cdms2010dark}. Direct detection experiments aim to measure energy depositions of order 100~keV and below, and suppress backgrounds to the level of less than one event per tonne per year to create a signal search region free from backgrounds.
 
DEAP-3600 (Dark matter Experiment using Argon Pulse-shape discrimination) has been designed to perform a direct WIMP dark matter search using 3600~kg of liquid argon (LAr) as a target. DEAP-3600 is located 2~km underground (6000~meters water equivalent overburden) at SNOLAB in Sudbury, Ontario, Canada~\cite{muonSNOLAB} and builds on the technology developed on the DEAP-1 prototype containing 7~kg of LAr~\cite{boulay2009measurement}. The projected sensitivity to the spin-independent WIMP-nucleon cross-section is $10^{-46}~\rm{cm}^{2}$ for a WIMP mass of 100~GeV/$c^2$~\cite{boulay2009measurement, deap50t}. Careful design and construction of the DEAP-3600 detector have reduced the predicted backgrounds in the search region of interest to less than 1~event in a fiducial exposure of 3~tonne-years.

Central features of the detector design include:
\begin {enumerate}
\item Single phase target: a monolithic inner volume of LAr allows minimal detector material to be in contact with the argon target, which can be made very radiopure. Particle energy deposition in the single phase LAr produces scintillation photons. The LAr scintillation time structure provides discrimination between WIMP-induced nuclear recoil events and electromagnetic background events~\cite{boulay2009measurement}. In addition to event characterization, event locations can be reconstructed using the detected scintillation signals.
\item Use of an acrylic cryostat: acrylic can be produced in a very controlled and radiopure fashion. As a hydrogenous material, it is a good neutron shield and possesses useful optical, mechanical, and thermal properties. The capacity for a large thermal gradient across acrylic allows for light guides of reasonable length with a cryogenic inside surface and an outside surface coupled to near-room-temperature photomultiplier tubes (PMTs) for signal readout. 
\item Radiopure raw materials: the detector is constructed of ultra-clean materials.  A quality assurance and testing program during procurement, manufacture, and construction provides careful control and inventory of radioactive contaminants. 
\item Electronics optimized for LAr scintillation detection: excellent single photoelectron response and digitization over 16~$\upmu$s enable full exploitation of the scintillation time structure for particle identification, with low deadtime and manageable data rates.
\end {enumerate}

\subsection{Detector Overview}
\begin{figure}[!p]
\centering\vspace*{-0.4in}
\includegraphics[width=4in, trim =  0 18mm 0 0, clip=true]{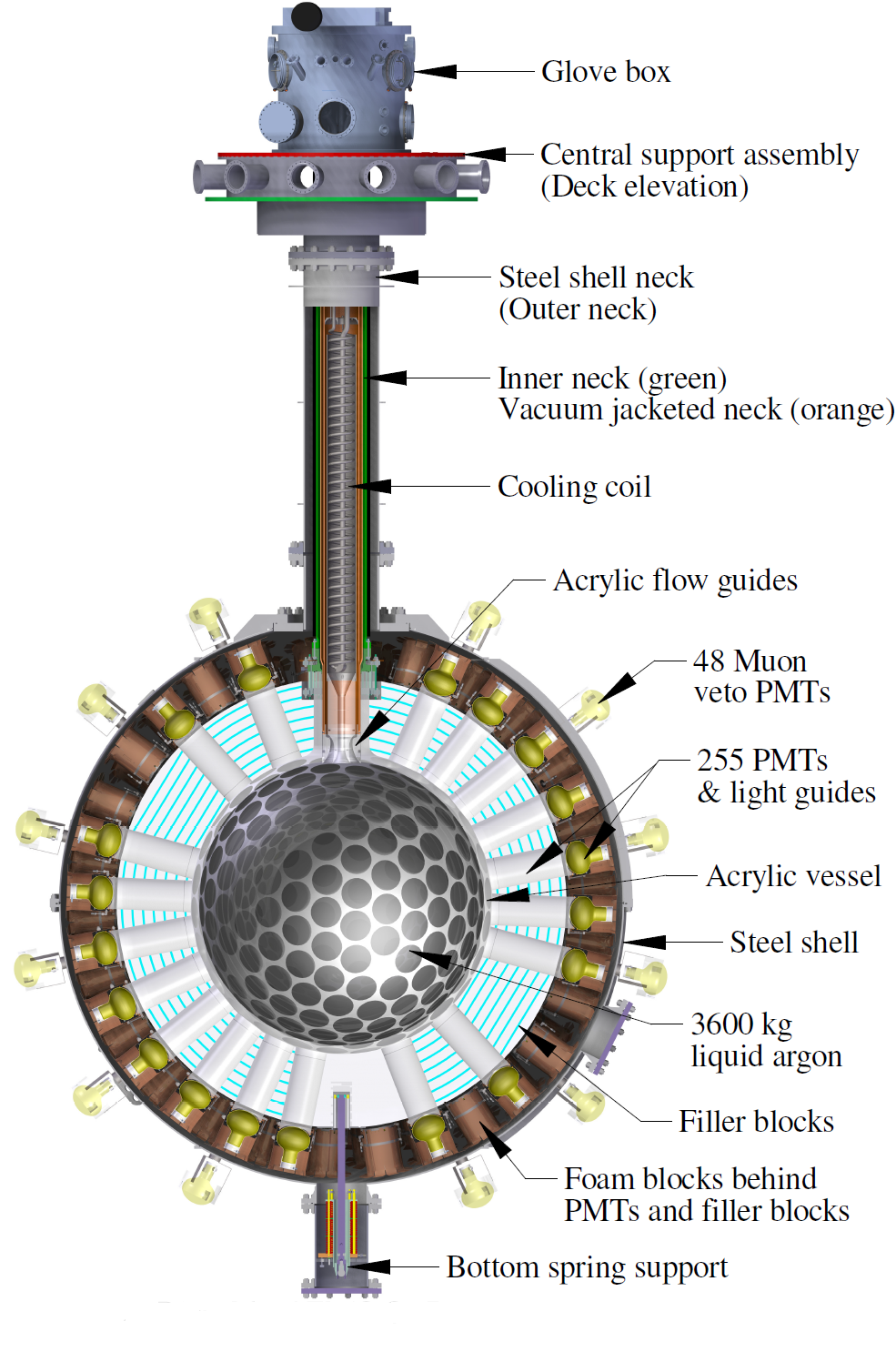}
\caption{The DEAP-3600 detector design showing the acrylic vessel, light guides, filler blocks, steel shell, neck, and glove box. Not shown are the wavelength shifting coating over the interior of the acrylic vessel and the surrounding muon veto water tank.}
\label{fig:design}
\end{figure}

A schematic view of the DEAP-3600 detector is shown in Figure~\ref{fig:design}, and the main design parameters are summarized in Table~\ref{tab:deap-3600parameters}. The inner detector includes the acrylic cryostat, neutron shielding materials, and the array of PMTs that view the LAr volume. The material selection and assay campaign for all detector materials are described in Section~\ref{sec:MatSelect}. The LAr purification system is described in Section~\ref{ssec:cryo} and inner detector components are described in Section~\ref{ssec:ID}.

The cryostat consists of a 5-cm-thick spherical acrylic vessel (AV), 85~cm in inner radius which can contain 3600~kg of LAr. Acrylic light guides (LGs), 45~cm in length, are directly bonded to the vessel. The LGs couple the AV to the PMTs and provide neutron shielding. Interspersed between the LGs are filler blocks comprised of layers of high-density polyethylene and polystyrene, which complete the neutron-shielding sphere. The inner surface of the AV is coated with a 3-$\upmu$m-thick layer of the organic wavelength-shifter 1,1,4,4-tetraphenyl-1,3-butadiene (TPB, $\rm C_{28}H_{22}$), deposited in situ~\cite{broerman2017application}, to convert the argon scintillation light into the visible wavelength region for transmission through the LGs to the PMT array. The target volume is viewed by 255~8-inch-diameter Hamamatsu R5912-HQE high quantum efficiency (HQE) PMTs manufactured with low radioactivity glass. The PMTs and associated readout electronics are described in Sections~\ref{sec:LightDetection} and~\ref{sec:electronics}, respectively.

The inner detector is housed in a stainless steel pressure vessel, comprising a spherical shell and the outer neck. Access to the inner detector volume is through a steel and acrylic neck that couples to the AV. This neck contains a cooling coil which uses liquid nitrogen (LN$_2$) to cool the LAr. A glove box interface at the top of the neck allows insertion or extraction of equipment in a radon-free environment. The outer steel shell is immersed in a 7.8-m-diameter shield tank filled with ultra-pure water and instrumented with 48~Hamamatsu R1408~PMTs serving as a muon veto. The pressure vessel and shield tank are described in Section~\ref{ssec:infrastructure}. Safety systems for maintaining a large, liquid argon volume in an underground environment are described in Section~\ref{sec:ODH}.

\begin{savenotes}
\begin{table}
\caption{DEAP-3600 detector design parameters for a 3-tonne-year exposure.}  
\centering
\begin{tabular}{@{}l  l@{}}
\toprule
Parameter & Design Specification  \\
\midrule
Sensitivity at 100~GeV/$c^2$ & 10$^{-46}$ cm$^2$  \\
Backgrounds target & $<0.6$~events   \\
Nominal region of interest (ROI) & 120--240~photoelectrons \\
Nominal analysis threshold & 15~keV$_{\rm ee}$ \\
Fiducial mass, radius & 1000~kg, 55~cm   \\
\midrule
Number of HQE inner detector PMTs & 255  \\
Light yield & 8~photoelectrons/keV$_{\rm ee}$ \\
Nominal position resolution at threshold & 10~cm\footnote{Updated simulations predict an improved position resolution.}\\
Total argon mass, radius & 3600~kg, 85~cm   \\
Water shielding tank diameter x height & 7.8~m x 7.8~m \\
Number of Cherenkov veto PMTs & 48 \\
\bottomrule
\end{tabular}
\label{tab:deap-3600parameters}
\end{table}
\end{savenotes}

A charged particle traversing the LAr target in DEAP-3600 loses energy through ionization. Scintillation emission is produced through the excitation or ionization of neutral argon atoms and the resulting creation of unstable argon dimers. These dimers decay with a characteristic singlet or triplet state lifetime of 6~ns or 1.5~$\upmu$s, respectively, and the ratio of production into these two states is dependent on the linear energy transfer of the incident radiation~\cite{doke2002absolute}. The scintillation efficiency for nuclear recoil-induced events is quenched by approximately 0.25 with respect to events induced by electron recoils~\cite{Gastler2012}. The resulting photon spectrum from argon dimer decay is in the vacuum ultraviolet (VUV) region, peaked at 128~nm~\cite{cheshnovsky1972emission}. This is lower in energy than the first atomic excited state of neutral argon, permitting the photons to travel through the argon without absorption. When these VUV photons reach the inner surface of the AV, they are absorbed by the TPB coating and re-emitted in the visible region, peaked at approximately 420~nm~\cite{gehman2011fluorescence}, near the peak quantum efficiency of the PMTs~\cite{ref:r5912}. The wavelength-shifted photons are transmitted through the acrylic light guides to the PMTs. The output signal is split into high- and low-gain channels, amplified, and shaped with custom electronics. 

\subsection{Design Realization}
The design required research and development in many areas of detector composition and construction. As this is the first use of a large acrylic cryostat, the mechanical properties of acrylic and acrylic bonds at liquid argon temperature (87~K) were measured at Los Alamos National Laboratory~\cite{LANLacrylicTest}. Large thermal stresses from differential contraction during operation and cooling were modeled with finite element analysis (FEA) and tested in detail. A technique for bonding the densely-packed LGs onto the acrylic sphere that is cryogenically and mechanically robust, clean, geometrically precise, and optically transparent was developed. A sanding resurfacer robot was built of radiopure materials with the ability to remove contamination from the entire inner acrylic surface. A method of large-scale, thin-film deposition, used to coat the inner acrylic surface with the wavelength shifter was developed and deployed~\cite{broerman2017application}. Lastly, an argon purification and radon-removal system was developed. 

The aim throughout design and construction was to minimize internal and external background interactions that could mimic a dark matter signal. These include: (i) backgrounds from electromagnetic events, particularly the beta decay of the naturally-occurring \Nuc{Ar}{39}; (ii) radon and radon progeny in the argon volume; (iii) radioactivity at or near the inner acrylic surface; (iv) neutrons; and (v) backgrounds associated with cosmic rays. 
 
Mitigation of the large~\Nuc{Ar}{39} background (approximately 1~Bq per kg of natural argon~\cite{benetti2007measurement}) is accomplished using pulse-shape discrimination (PSD) of the scintillation signal. This technique is very powerful in LAr and has a projected discrimination power of $10^{-10}$~\cite{boulay2009measurement, Boulay2006} for DEAP-3600. The HQE PMTs are operated near room temperature to allow optimal performance and light collection, upon which the PSD strongly depends.

The radon background mitigation strategy employs material Rn emanation assay and selection, control of exposure to lab air during production, and surface treatment after construction, in addition to the cryogenic purification of the argon. Radon (\Nuc{Rn}{222} and~\Nuc{Rn}{220}) and progeny in the LAr target target itself are reduced by carefully controlling the materials and construction of the argon purification systems and inner detector. 

The acrylic for the cryostat was chosen carefully, and its production was monitored directly by the DEAP-3600~collaboration, minimizing exposure of the acrylic precursors to radon and other contaminants. Construction of the acrylic vessel was also highly controlled; however, the final assembly steps were performed underground at SNOLAB, where the~\Nuc{Rn}{222} level is approximately 130~Bq/m$^3$. The robotic resurfacer was built to remove radio-contaminants built up on the inner acrylic surface from radon surface deposition and diffusion following the AV construction underground. To minimize the contribution to surface backgrounds from the wavelength shifting coating applied after resurfacing, the production of the TPB was monitored by the collaboration. Finally, surface backgrounds are reduced in analysis by using position reconstruction algorithms to reject events near the inner AV surface. 

Neutrons produced internally in the detector are controlled with radiopure materials to reduce their production rate and with hydrogenous shielding to thermalize those that are produced near the LAr. The primary source of neutrons, the borosilicate PMT glass, is effectively moderated in the acrylic LGs and polyethylene filler blocks. Neutrons from the rock wall at SNOLAB are moderated by the water-filled shield tank. Cosmic ray muons, with a flux of 0.27~m$^{-2}$ day$^{-1}$~\cite{muonSNOLAB}, are tagged with the Cherenkov light detection veto system to reject the cosmogenic neutrons they produce. 

Table~\ref{tab:backgroundBudget} summarizes the background goals for a 3~tonne-year fiducial exposure in the energy ROI of 120--240 photoelectrons, corresponding to a nominal energy ROI of 15~to 30~keV$_{\rm{ee}}$.

\begin{table}[h!]
\caption{Targeted number of events in the energy ROI, $120<$~photoelectrons~$<240$, with a 3-tonne-year fiducial exposure. Fiducialisation assumes both a 10~cm position resolution for surface events and 50\% nuclear recoil acceptance from pulse shape discrimination.}
\begin{center}
\begin{tabular}{@{}l  l  l@{}}
\toprule
Source & Events in Energy ROI & \parbox[c]{2.75 cm}{\centering Fiducial Events in Energy ROI}\\ 
\midrule
Neutrons & 30 & $<0.2$ \\ 
Alphas (surface) & 150 & $<0.2$ \\ 
Betas/gammas (\Nuc{Ar}{39} dominated) & $1.6\times 10^{9}$ & $<0.2$ \\
\midrule 
Sum &  & $<0.6$ \\ 
\bottomrule
\end{tabular}
\end{center}
\label{tab:backgroundBudget}
\end{table}
\section{Material Selection}\label{sec:MatSelect}
Detector materials were carefully selected for radiopurity and optical properties to maximize the dark matter detection sensitivity. An extensive radiopurity assay campaign was performed in combination with Monte Carlo simulations to ensure adherence to the targeted background budget. 

\subsection{Detector Simulation} \label{subsection:DetectorSimulation}
The DEAP-3600 detector simulation uses RAT~\cite{tcald_aarm_2014}, a software framework for simulation and analysis of liquid scintillator experiments, which uses Geant4~\cite{Agostinelli2003250} version 4.9.6 and ROOT~\cite{brun1997root} version 5.34 libraries. Customized versions of RAT are currently used by the MiniCLEAN, SNO+, and DEAP-3600 collaborations. 

Simulations have been used extensively for the definition of radiopurity requirements for detector materials, background rejection studies including shielding optimization and position reconstruction, light yield optimization, activity requirements for calibration sources, and studies of electronics or trigger-related biases and other systematic effects.

The simulation implements the full, as-built detector geometry, including the SNOLAB Cube Hall cavern in which the DEAP-3600 detector is located. A number of optical parameters of the simulation are defined by ex-situ measurements, which include the wavelength-dependent light attenuation length of acrylic~\cite{bodmer2014measurement}, wavelength-dependent reflectance of the diffuse\footnote{Data kindly contributed by Martin Janecek (LBNL), as measured with the apparatus described in~\cite{Janecek}.} and specular reflectors surrounding the AV, as well as the alpha scintillation properties of TPB, including light yield~\cite{Pollmann2011}, the scintillation time profile information, and its temperature dependence~\cite{LaurelleMSc, veloce2016temperature}. Optical photons are fully propagated in RAT and the process has been validated using DEAP-1 data~\cite{Amaudruz2014}. 

The following extension packages are used in the simulation in addition to the standard Geant4 physics processes: 
\begin{itemize}
	\item A detailed simulation model for nuclear recoils on arbitrarily rough surfaces was developed~\cite{Kuzniak2012} to investigate backgrounds originating from the inner surface of the AV. The model combines explicit surface roughness implementation in the geometry with a Geant4 extension, available as one of its extended examples (TestEm7), which contains all physics relevant for multiple inter-atomic and alpha scattering in the 10~keV--10~MeV energy range~\cite{Screen, G4Screen}. It has been extensively benchmarked against SRIM~\cite{ziegler2010srim} with respect to nuclear straggling and implantation, as well as backscattering. 
	\item Argon scintillation is simulated based on the model from~\cite{Mei200812}. The NEST (Noble Element Scintillation Technique) model~\cite{NEST}, developed primarily for xenon and used by LUX, EXO, and XENON100 collaborations, is optionally available in RAT. SCENE measurements to calculate nuclear recoil quenching factors and PSD distributions are also implemented in RAT.
	\item The hadronic physics models relevant for muon- and gamma-induced neutron simulations recommended in~\cite{Lindote2009366, HornPhD} have been adopted. At energies below 20~MeV, high precision data-driven neutron models provided by Geant4 are used.
	\item The energy spectrum and rate of neutrons from inner detector components is calculated using the SOURCES-4C code~\cite{Wilson2009608} and cross-checked using NeuCBOT~\cite{westerdale2017radiogenic}.
\end{itemize}

Based on the background targets in Table~\ref{tab:backgroundBudget}, material selection efforts were focused on mitigating backgrounds from alphas and neutrons. The activity from \Nuc{Ar}{39} beta decays dominates the overall rate unless special, sequestered argon sources depleted in cosmogenically-produced \Nuc{Ar}{39} are employed~\cite{xu2012study, agnes2016results}. These events, however, are readily removed from the analysis using PSD. Each potential contributing source to the alpha and neutron backgrounds is normalized to a total of 0.2 background events in a 3-tonne-year fiducial exposure to set target values.

In the background model, the leading alpha backgrounds originate from contamination in the wavelength shifter and acrylic bulk. While traversing these regions, alphas can lose energy, reducing the typically high energy signal down into the WIMP search ROI. The targeted alpha background activity is set by normalizing to 0.2~background events in a 3-tonne-year fiducial exposure and assuming a 10~cm position reconstruction resolution at threshold, resulting in a factor of $10^{3}$ reduction in surface background rate in analysis. The targeted activity of the \Nuc{U}{238}, \Nuc{Th}{232}, and \Nuc{Pb}{210} chains for TPB and acrylic is summarized in Table~\ref{tab:alphaTarget} assuming secular equilibrium and 1~alpha from \Nuc{Pb}{210} per \Nuc{U}{238} decay. 

\begin{table}[h!]
\caption{Targeted activity of the \Nuc{U}{238}, \Nuc{Th}{232}, and \Nuc{Pb}{210} decay chains in TPB and AV acrylic contributing to the surface alpha background assuming secular equilibrium apart from \Nuc{Pb}{210}. Surface alphas can lose energy in the acrylic and TPB layer reducing the energy deposited in the LAr volume into the WIMP search ROI. The sources of alphas are normalized to a total of 0.2~background events in a 3-tonne-year fiducial exposure to report the targeted values.}
\begin{center}
\begin{tabular}{@{}l l l l@{}}
\toprule
 & \multicolumn{3}{r@{}}{Targeted Activity [$\upmu$Bq/kg]} \\
Component & \ \ \Nuc{U}{238} \ \ \ & \Nuc{Th}{232} \ \ \ & \Nuc{Pb}{210} \ \ \ \\ 
\midrule
TPB			& \ \ 5.7 	& \ \ 8.8	& \ 4.0  \\ 
AV acrylic bulk & \ \ 2.9 	& \ \ 3.9	& \ 20.0 \\ 
\bottomrule
\end{tabular}
\end{center}
\label{tab:alphaTarget}
\end{table}

The main sources of internal neutron backgrounds come from the production of neutrons through $(\alpha,n)$ reactions and spontaneous fission in the detector materials. The dominant components contributing to the neutron backgrounds in the simulation model are summarized in Table~\ref{tab:neutronTarget}, along with the targeted radiopurity values reported by normalizing to a total 0.2~background events from neutrons in a 3-tonne-year fiducial exposure.

\begin{table}[h]
\caption{Specific activity targets of the \Nuc{U}{238}, \Nuc{U}{235}, and \Nuc{Th}{232} decay chains in detector materials contributing to the neutron background assuming secular equilibrium. Target values are reported by normalizing to a total of 0.2~background events in the WIMP energy ROI with a 3-tonne-year fiducial exposure from neutrons emitted by $(\alpha,n)$ and spontaneous fission reactions.}
\begin{center}
\begin{tabular}{@{}lclcl@{}}
\toprule
& \multicolumn{3}{c}{Targeted Activity [mBq/kg]} \\
Component 	& \Nuc{U}{238} & \Nuc{U}{235} & \Nuc{Th}{232} \\
\midrule
PMT glass	& ~~~~82.8~~~~ & 72.0 & 47.2 \\
PMT ceramic	& ~~~3530~~~ & --- & 960 \\
AV acrylic		& 0.02 & 0.09 & 0.08 \\
LG acrylic 	& 0.12 & 0.19 & 0.16 \\
Filler blocks (polyethylene)	& 0.36 & 0.53 & 0.54 \\
PMT mount PVC	& 124.0 & 72.0 & 49.2 \\
Neck Steel 	& 19.2 & 96.0 & 19.2 \\
Neck PMT glass 	& 24300 & --- & 11600 \\
\bottomrule
\end{tabular}
\end{center}
\label{tab:neutronTarget}
\end{table}%

The radon emanation rate for argon-wetted materials (purification system, detector inner-neck components) is constrained by the allowable alpha background on the inner AV source. For the region outside the AV but within the steel shell, the allowable radon load is determined by assuming that progeny collect on the outer AV surface approximately 5~cm from the inner AV surface, using the conservative assumption that all progeny will stick to the detector surfaces. The targeted radon emanation rates for inner detector components to maintain the neutron background target are shown in Table~\ref{tab:rnTarget}. In practice during detector operation, the steel shell region is purged with radon-scrubbed boil-off nitrogen gas; the actual radon load in this region is should be that of the purge gas. 

\begin{table}[h!]
\caption{Targeted radon emanation rates for major inner detector materials and components. 
Target values are reported by normalizing to a total of 0.2~background events in the WIMP energy ROI with a 3-tonne-year fiducial exposure from neutrons emitted by $(\alpha,n)$ reactions by radon plated-out on cold inner-detector surfaces.
}
\begin{center}
\begin{tabular}{@{}l l@{}}
\toprule
Radon Emanation Source  		& Target Emanation Rate  \\ 
\midrule
PMT cables 					& 0.047 [mBq/m] \\ 
AV acrylic 					& 5.5 [mBq/m$^2$]  \\
LG acrylic						& 1.2 [mBq/m$^2$]  \\ 
Filler blocks 					& 0.7 [mBq/m$^2$] \\ 
PMT mount (PVC) 				& 1.8 [mBq/m$^2$]  \\ 
FINEMET PMT magnetic shielding~\cite{ref:finemet}		& 2.0 [mBq/m$^2$] \\ 
Stainless steel shell 				& 2.7 [mBq/m$^2$] \\ 
PMTs 						& 0.4 [mBq/PMT] \\ 
\bottomrule
\end{tabular}
\end{center}
\label{tab:rnTarget}
\end{table}%

\subsection{Material Assay Techniques}
\label{ssec:assay}
To reach the radiopurity goals summarized in Tables~\ref{tab:alphaTarget},~\ref{tab:neutronTarget}, and~\ref{tab:rnTarget}, extensive low-background gamma assay and radon emanation measurement programs to select materials, in addition to material and handling quality assurance programs, were developed.

SNOLAB has a well-established gamma assay program; a 200 cm${^3}$ high-purity germanium well detector (Princeton Gamma-Tech Instruments, Inc.)~\cite{lawson2016ultra} was purchased and installed at SNOLAB to meet the assay requirements for the DEAP experiment. An inventory of gamma assay measurements\footnote{Radiopurity database: \url{https://deap-radiopurity.physics.carleton.ca/database/}} for the \Nuc{U}{238}, \Nuc{Th}{232}, and \Nuc{U}{235} decay chains which contribute to the neutron background from major detector components is listed in Table~\ref{tab:gammaResults} and from tools used in manufacturing components in Table~\ref{tab:gammaResults2}. Material assays measuring the \Nuc{Th}{234} and \Nuc{Pa}{234m} gamma lines in the \Nuc{U}{238} decay chain often show different activities from what is observed in the \Nuc{Ra}{226}, \Nuc{Pb}{214}, and \Nuc{Bi}{214} gamma lines. This discrepancy indicates that secular equilibrium is often broken between \Nuc{Th}{230} and \Nuc{Ra}{226}. \Nuc{U}{238} and its progeny up to and including \Nuc{Th}{230} is referred to as the ``{\Nuc{U}{238}$_\text{upper}$~chain", while \Nuc{Ra}{226} and its progeny as the ``{\Nuc{U}{238}$_\text{lower}$}~chain'', assuming secular equilibrium within each sub-chain. Assay results contributing to the electromagnetic backgrounds (\Nuc{K}{40}, \Nuc{Co}{60}) are measured, but not reported in Table~\ref{tab:gammaResults}. Uncertainties arise from counting statistics and detection efficiency. A description of the components can be found in Sections~\ref{ssec:ID} and~\ref{sec:LightDetection}.

\begin{sidewaystable}
\caption{Gamma assay results for major detector components. A description of the components can be found in Sections~\ref{ssec:ID} and~\ref{sec:LightDetection}. Activities are reported with 1-sigma uncertainties. A 90\% confidence limit is placed when the measurement is below the background sensitivity of the detector. It is assumed that secular equilibrium is broken between \Nuc{Th}{230} and \Nuc{Ra}{226} in the \Nuc{U}{238} decay chain.}
\label{tab:gammaResults}
\centering
\begin{tabular}{@{}lllll@{}}
\toprule
Component						& {\Nuc{U}{238}$_\text{lower}$} 			 	& {\Nuc{U}{238}$_\text{upper}$}				 &{\Nuc{Th}{232}}			& {\Nuc{U}{235}} 		\\
& \multicolumn{4}{c}{\hspace{-2mm} [mBq/kg]} 					\\
\midrule
Methyl methacrylate monomer (LG bonding)	&		$1.4~\pm1.0$					& $<15$								& $<0.9	$ 				& $<1.8$				\\
AV acrylic							&		$<0.1$						& $<2.2$								& $<0.5$					& $<0.2$				\\
Acrylic beads (RPT)					&		$<3.1$						& $16\pm15$							& $0.8\pm0.3$				& $0.6\pm0.5$			\\ 		
LG acrylic                   					&		$<0.1$						& $<9.0$          							& $<0.3$ 					& $<0.6$				\\
304 welded stainless steel (steel shell)	&		$1.4\pm1.1$					& $<5.0$								& $4.7\pm1.5$				& $<3.3$				\\
304 stainless steel stock (steel shell)	&		$2.1\pm1.1$					& $<112$								& $1.9\pm1.1$				& $<5.4$				\\
316 stainless steel bolts (steel shell) 	&		$<6.1$						& $<315$								& $94\pm9$				& $<17$				\\
Carbon steel (stock)					&		$2.0\pm0.7$ 					& $111\pm43$							& $10.0\pm1.0$			& $8.6\pm1.9$			\\
Invar steel (neck)					&		$4.5\pm1.5$					& $120\pm77$							& $2.5\pm1.5$				& $<3.6$				\\
R5912 HQE PMT glass				&	       	$921\pm34$					& $225\pm114$					   	& $139\pm7$				& $25\pm3$			\\
R5912 HQE PMT ceramic				& 		$978\pm56$					& $15500\pm2800$						& $245\pm28$				& $503\pm51$			\\
R5912 HQE PMT feedthrough pieces	&		$1140\pm60$					& $2350\pm1460$						& $430\pm32$				& $38\pm9$			\\
R5912 HQE PMT metal components	& 		$<5.5$						& $-$								& $<3.3$					& $-$				\\
RG59 PMT cable (Belden E82241) 		&		$4.5\pm1.3$  					& $91\pm46$  							& $1.2\pm0.9$				& $3.4\pm1.4$			\\
PMT mount PVC (Harvel)				&		$72\pm5$						& $232\pm130$					   	& $18.6\pm2.5$			& $5.6\pm1.5$	 		\\
PMT mount copper					&		$<0.5$						& $<10$								& $<0.8$					& $<1.3$				\\
Neck Veto PMT glass				&		$1230\pm620$					& $-$								& $407\pm203$			& $57\pm29$				\\
Filler block polyethylene 				&		$0.4\pm0.3$					& $<14$								& $<0.1$					& $<0.15$				\\
Filler block Styrofoam~\cite{ref:styrofoam} &		$33.5\pm3.4$ 					& $115\pm64$							& $<1.5$					& $<1.4$				\\
White Tyvek paper (diffuse reflector)		&		$<0.3$						& $50\pm37$							& $1.3\pm0.8$				& $<2.2$				\\
Black Tyvek paper (LG wrapping)		&		$<1.8$						& $<127$								& $5.6\pm2.3$				& $<3.8$	 			\\
Black polyethylene tube (upper neck)	&		$13.7\pm1.8$					& $<60$								& $3.2\pm1.1$				& $2.6\pm1.4$			\\
TPB (Sigma Aldrich)					&		$< 3.9$						& $-$								& $<8.7$					& $-$ 				\\
\bottomrule
\end{tabular}
\end{sidewaystable}

\begin{sidewaystable}
\caption{Gamma assay results for tooling used during construction and manufacture of detector components. Activities are reported with 1-sigma uncertainties. A 90\% confidence limit is placed when the measurement is below the background sensitivity of the detector. It is assumed that secular equilibrium is broken between \Nuc{Th}{230} and \Nuc{Ra}{226} in the \Nuc{U}{238} decay chain.}
\label{tab:gammaResults2}
\centering
\begin{tabular}{@{}lllll@{}}
\toprule
Component														& {\Nuc{U}{238}$_\text{lower}$}  			& {\Nuc{U}{238}$_\text{upper}$}		&{\Nuc{Th}{232}}			& {\Nuc{U}{235}} \\
& \multicolumn{4}{c}{\hspace{-2mm} [mBq/kg]} 							\\
\midrule
\multicolumn{5}{l}{\hspace{-2mm}$\mathit{Purification~System~Welding}$} 	\\
TIG weld sample	 												& $7.7\pm5.7$							& $<27$						& $25.2\pm7.8$ 			& $<16$	\\
SMAW weld sample  												& $<23$		 						& $<1255$					& $51.9\pm12.2$ 			& $<13$	\\	
Welding electrodes A (Blue Demon TE2C-116-10T)							& $221\pm65$							& $<493$						& $1890\pm184$			& $<56$	\\
Welding electrodes B (Blue Demon TE2C-116-10T)							& $66.6\pm42.6$						& $<1300$					& $710\pm103$			& $<138$	\\
Welding electrodes C (Blue Demon TE2C-116-10T)							& $86.1\pm21.8$						& $<642$						& $911\pm73$				& $<108$	\\
Weld filler rods														& $<4.8$								& $<157$						& $3.0\pm2.5$				& $<1.8$	\\		
\midrule
\multicolumn{5}{l}{\hspace{-2mm}$\mathit{Inner~AV~Sanding}$} 			\\
Brazed diamond sanding pad (Superabrasives)							& $141\pm24$							& $<845$						& $49.8\pm17.9$			& $31\pm19$\\ 
Plated diamond sanding pad (Superabrasives)							& $4680\pm283$						& $<4130$					& $6180\pm300$			& $218\pm64$\\
3M 6002J flexible diamond pads										& $25.1\pm15.4$						& $<785$						& $<10.8$					& $<33$	\\
Diamond sandpaper	 (Diamante Italia)									& $3120\pm136$						& $<2300$					& $3370\pm125$			& $157\pm22$\\
Red sandpaper (RPT)												& $48.7\pm19.7$						& $<335$						& $< 10.1$				& $<32$\\
\midrule
\multicolumn{5}{l}{\hspace{-2mm}$\mathit{LG~Acrylic~Polishing}$} 			\\
Diamond lapping film (3M 661X)										& $142\pm38$							& $<882$						& $93.6\pm35.0$			& $<31$	\\
Diamond lapping film (3M 661X)	 									& $94.0\pm16.5$						& $<276$						& $105.\pm18.1$			& $<33$\\
\bottomrule
\end{tabular}
\end{sidewaystable}

A new radon emanation measurement system constructed at Queen's University, similar to that described in~\cite{liu1993222rn}, was used to qualify and select detector materials. Materials are loaded into a vacuum chamber, the chamber is evacuated, and the material is then allowed to emanate into vacuum before the Rn atoms are collected using a cold trap. The alpha decay rate from the Rn atoms is measured to determine the initial radon emanation rate from the material. Uncertainties in the emanation rate are due to system backgrounds, and the efficiencies of trapping and detection. Typical backgrounds in the emanation system are on the order of a single radon atom. Upper limits are set when there is no signal above background. Table~\ref{tab:rnResult} lists results from the \Nuc{Rn}{222} emanation of the main detector components and tooling used during fabrication. Limits on the AV and LG acrylic emanation are both set at $<0.3$~mBq/m${}^2$; the AV radon emanation is based on the assayed uranium content and not a direct radon emanation measurement. The steel shell volume is purged using a boil-off nitrogen gas system with a radon emanation rate of 5 $\upmu$Bq/kg~\cite{simgen2006ultrapure}.

\begin{savenotes}
\begin{table}[h!]
\caption{Measured \Nuc{Rn}{222} emanation rates for components used in the DEAP-3600 detector contained within the stainless steel shell. A description of the most important components can be found in Section~\ref{ssec:ID}. Uncertainties arise from counting, sample emanation times, and detection and trapping efficiency.}
\begin{center}
\begin{tabular}{@{}l l@{}}
\toprule
Source 									& Emanation Rate\\
\midrule
										& [mBq/m$^{2}$] \\
Filler blocks 								& $1.6\pm0.5$\\
FINEMET PMT magnetic shielding~\cite{ref:finemet}		& $0.8\pm0.2$ \\
ESR film reflector~\footnote{Candidate material, not used in final construction.} 	& $<2.2$  \\
Tyvek diffuse reflector 						& $<0.1$ \\
Black tyvek absorber						& $0.4\pm0.2$ \\
PMT mount PVC (McMaster-Carr stock)			& $<0.7$\\ 
PMT polyethylene foam						& $<0.9$\\
Teflon sheets (McMaster-Carr stock)				& $0.4\pm0.2$ \\ 
High density polyethylene pipe	 				& $3.5\pm0.8$ \\
304 Stainless Steel (McMaster-Carr stock)		& $<1.6$ \\ 
Carbon steel (McMaster-Carr stock)				& $0.6\pm0.1$\\
White PMT mount adhesive styrofoam sheet		& $<1.5$ \\
Stycast 1266 A/B (Emerson \& Cuming)			& $<4.2$ \\
\midrule
										& [mBq/m] \\
RG59 PMT cable (Belden E82241)				& $0.026\pm0.001$\\
Steel shell EPDM O-ring	 					& $16.1\pm1.8$ \\
Viton O-ring 								& $1.3\pm0.2$ \\ 
Buna 451 O-ring 							& $17\pm2$\\ 
\midrule
										& [mBq/unit] \\
Hamamatsu R5912 PMTs 					& $< 0.3$  \\ 
PMT mount O-ring 							& $0.3\pm0.1$\\
\bottomrule
\end{tabular}
\end{center}
\label{tab:rnResult}
\end{table}
\end{savenotes}

Radon emanated from the resurfacer sanding robot, which contributes to the surface alpha background through possible collection of the long-lived radon daughter \Nuc{Pb}{210} while the sanding robot is deployed in the AV, was calculated based on screening measurements of the individual components to be less than 20~mBq.

The \Nuc{Pb}{210} targets set by the maximum allowable contribution to $(\alpha,n)$ backgrounds are below the sensitivity of most current assay techniques. A program based on vaporization and subsequent chemical processing originally developed by the Sudbury Neutrino Observatory (SNO)~\cite{NantaisThesis} was extended to allow a sensitive assay of \Nuc{Pb}{210} in acrylic. Samples of the acrylic were vaporized, the residue extracted by rinsing with an acidic Aqua Regia solution, and the effluent collected and counted in the germanium well detector at SNOLAB. Details of the DEAP-3600 acrylic vaporization process can be found in~\cite{NantaisThesis}. In addition, the 5.3~MeV alphas from the decay of \Nuc{Po}{210} were counted from the acid solution by plating it out on nickel discs, with a 450~mm$^2$ ORTEC ULTRA-AS ion-implanted-silicon detector~\cite{NantaisThesis}. From these measurements, which affects the \Nuc{Pb}{210} surface alpha background, an upper limit of 0.62~mBq/kg \Nuc{Pb}{210} was set for the AV acrylic. 

Based on results from the extensive material assay campaign shown in Tables~\ref{tab:gammaResults},~\ref{tab:gammaResults2}, and~\ref{tab:rnResult}, the expected number of background events in a 3-tonne-year fiducial exposure is assessed. The expected number of alpha background events assuming a 10~cm position resolution is shown in Table~\ref{tab:alphaExpectation}. The sensitivity of the TPB assay given the amount of product available was not sufficient to test adherence to the required alpha background. A number of steps were undertaken during the synthesis of the TPB in coordination with the manufacturer and during storage and deposition to control possible contaminations. The expected number of background events from neutrons produced in $(\alpha,n)$ reactions and spontaneous fission is shown in Table~\ref{tab:neutronExpectation}, whereas neutrons produced from $(\alpha,n)$ reactions and spontaneous fission due from radon plate-out on cold inner-detector surfaces is shown in Table~\ref{tab:rnExpectation}. 

\begin{table}[h!]
\caption{Expected number of alpha background events based on screening measurements. Event numbers are given for the WIMP energy ROI with a 3-tonne-year fiducial exposure assuming a 10~cm position resolution.}
\begin{center}
\begin{tabular}{@{}l l l l@{}}
\toprule
Component 	& \ \ \Nuc{U}{238} \ \ \ & \Nuc{Th}{232} \ \ \ & \Nuc{Pb}{210} \ \ \ \\ 
\midrule
TPB			& \ \ $<6.8$ 	& \ \ $<9.9$	& $-$  \\
AV acrylic bulk 	& \ \ $<0.3$ 	& \ \ $<1.3$	& $<0.3$ \\ 
\midrule
Total Events: $<18.7$ 	& \ \ $<7.2$ 	& \ \ $<11.2$	& $<0.3$   \\ 
\bottomrule
\end{tabular}
\end{center}
\label{tab:alphaExpectation}
\end{table}

\begin{sidewaystable}
\caption{Expected number of neutron background events from $(\alpha,n)$ reactions and spontaneous fission based on screening measurements. Event numbers are given for the WIMP energy ROI with a 3-tonne-year fiducial exposure assuming a 10~cm position resolution.}
\begin{center}
\begin{tabular}{@{}lllll@{}}
\toprule
Component 				& \Nuc{U}{238} 				& \Nuc{U}{235} 			& \Nuc{Th}{232} \\
\midrule
PMT glass				& $(11.1\pm0.4) \cdot 10^{-2}$   	& $(3.5\pm0.4) \cdot 10^{-3}$ 	& $(2.9\pm0.2) \cdot 10^{-2}$ \\
PMT ceramic				& $(2.8\pm0.2) \cdot 10^{-3}$   	        & -  	                                          & $(2.6\pm0.3) \cdot 10^{-3}$  \\
AV acrylic					& $< 5 \cdot 10^{-2}$ 			& $< 2 \cdot 10^{-2}$ 		& $< 6 \cdot 10^{-2}$ \\
LG acrylic 				& $< 8 \cdot 10^{-3}$  			& $< 3 \cdot 10^{-2}$ 		& $< 2 \cdot 10^{-2}$ \\
Filler blocks (polyethylene)	& $(1.1\pm0.8) \cdot 10^{-2}$ 		& $< 3 \cdot 10^{-3}$ 		& $< 2 \cdot 10^{-3}$ \\
PMT mount (PVC)			& $(5.8\pm0.4) \cdot 10^{-3}$ 		& $(8.0\pm2.0) \cdot 10^{-4}$ 	& $(3.8\pm0.5) \cdot 10^{-3}$ \\
Neck steel 				& $(1.1\pm0.6) \cdot 10^{-3}$  		& $<6 \cdot 10^{-4}$ 		        & $(1.0\pm0.6) \cdot 10^{-3}$  \\
Neck Veto PMT glass 			& $(5\pm3) \cdot 10^{-4}$   	        & -  	                                          & $(4\pm2) \cdot 10^{-4}$ \\
\midrule
Total Events: 0.23         				& $0.08$  						& $0.06$  					& $0.09$   \\
\bottomrule
\end{tabular}
\end{center}
\label{tab:neutronExpectation}
\end{sidewaystable}

\begin{table}[h!]
\caption{Expected number of neutron background events based on radon emanation measurement assuming $(\alpha,n)$ reactions and spontaneous fission induced from radon plate-out on cold inner-detector surfaces. Event numbers are given for the WIMP energy ROI with a 3-tonne-year fiducial exposure assuming a 10~cm position resolution.}
\begin{center}
\begin{tabular}{@{}l l@{}}
\toprule
Radon Emanation Source  		& Events  \\ 
\midrule
PMT cables 					& $(5.5\pm0.2) \cdot 10^{-3}$ \\ 
AV acrylic 					& $<5\cdot 10^{-4}$  \\
LG acrylic						& $<3\cdot 10^{-3}$ \\ 
Filler blocks 					& $(2.3\pm0.7) \cdot 10^{-2}$ \\ 
PMT mount (PVC) 				& $<4\cdot 10^{-3}$ \\ 
FINEMET PMT magnetic shielding~\cite{ref:finemet}		& $(4\pm1) \cdot 10^{-3}$  \\ 
Stainless steel shell 				& $<6\cdot 10^{-3}$  \\ 
PMTs 						& $<8\cdot 10^{-3}$  \\ 
\midrule
Total	Events						& $0.053$ \\
\bottomrule
\end{tabular}
\end{center}
\label{tab:rnExpectation}
\end{table}

\subsection{Material Production and Quality Assurance}\label{subsection:MaterialProduction}
Acrylic, or poly(methyl methacrylate) (PMMA), is a polymer of methyl methacrylate (MMA). Alpha decays on the inner surface of the acrylic can produce nuclear recoils in the LAr, while those that decay in the bulk acrylic may undergo the \Nuc{C}{13}$(\alpha,n)$\Nuc{O}{16} reaction~\cite{araki2005measurement} to produce neutrons, which can then scatter in the LAr to produce a nuclear recoil. The acrylic annealing process employed during construction to improve light collection also contributes to the background from radon daughters due to the temperature-dependence of radon diffusion in acrylic. These acrylic-related alpha and neutron backgrounds are mitigated by limiting radon exposure and controlling cleanliness during construction. 

\subsubsection{Cryostat Acrylic}
Reynolds Polymer Technologies, Inc. (RPT) Asia Ltd. (Rayong, Thailand) acquired the MMA from the Thai MMA Co.~plant in eastern Thailand~\cite{jillings2013control}. For DEAP-3600, the raw MMA was purchased directly from the production pipeline and delivered to RPT Asia, limiting excessive exposure to radon-laden air in storage tanks. 

To calculate the radon load during manufacture the contamination when MMA enters a production vessel was derived, assuming all radon in the volume was dissolved in MMA and any radon progeny in the volume also becomes trapped. The resulting \Nuc{Pb}{210} concentration in acrylic, using a density of MMA of approximately 1000~kg/m$^3$, is 
$\mbox{A}_{\mbox{\small acrylic}}(\Nuc{Pb}{210})$:
\begin{equation}\nonumber
\mbox{A}_{\mbox{\small acrylic}}(\Nuc{Pb}{210})[\mbox{mBq/tonne}] =
\frac{\lambda_{^{210}Pb}}{\lambda_{^{222}Rn}} \mbox{A}_{\mbox{\small air}}(\Nuc{Rn}{222})[\mbox{mBq/m}^{3}]
\end{equation}
where $\lambda$ is the decay constant for \Nuc{Pb}{210} or \Nuc{Rn}{222} and A$_\mathrm{air}$ is the activity of \Nuc{Rn}{222} in the environmental air. The radon levels at the MMA production plant were measured with a Durridge Rad-7 electronic radon detector~\cite{ref:rad7} to be consistent with sea-level concentrations of 1~Bq/m$^3$.  The solubility of radon in solid acrylic, 8.2~\cite{wojcik1991measurement}, is used to estimate background levels after the panels have been formed. 

Following all steps in production, the predicted \Nuc{Pb}{210} activity in the acrylic panels of the AV is 0.021~mBq/kg~\cite{jillings2013control}, which is negligible compared to the measured \Nuc{Pb}{210} upper limit in Section~\ref{ssec:assay}.

\subsubsection{Light Guide Acrylic}
The radiopurity requirements are less stringent for the LG acrylic as it is not in direct contact with the LAr. Selection of the LG acrylic was based on minimizing light attenuation. A wavelength-dependent attenuation length was calculated from transmission measurements of 10~different length samples ranging from 4~mm to 110~mm in a PerkinElmer Lambda 35~UV/Vis optical spectrometer. Acrylic with a UV absorbing additive was used for both the LGs and AV to minimize contributions from Cherenkov light generation in the acrylic. 

The attenuation length in acrylic can reach a few tens of meters for blue light. After receiving and qualifying the LG acrylic material from the supplier, more Rayleigh scattering was observed than had been present in earlier test samples. This additional Rayleigh scattering could be reduced by annealing the LGs near 85${}^{\circ}$C. This procedure increased the transmission at 440~nm, the mean TPB emission wavelength, by approximately a factor of two. Out of 10~acrylic suppliers, Spartech Polycast acrylic, with an attenuation length at 440~nm after annealing of $6.2\pm0.6$~m was selected for the LGs. The final acrylic delivered by Reynolds Polymer for the AV had an attenuation greater than 7~m above 420~nm. 

\subsubsection{TPB Wavelength Shifter}
The TPB required to cover the 9~m$^2$ inner surface of the AV must be radiopure to prevent alpha decays from the bulk of the TPB layer from producing background events in the region of interest. For this reason, special arrangements were made with the manufacturer, Alfa Aesar (Heysham, UK), in the synthesis of the TPB to meet the required radiopurity requirements shown in Table~\ref{tab:alphaTarget}. Base chemicals with assay certificates of 99\% purity or greater were used during production, and all steps in the synthesis process were performed under a boil-off nitrogen atmosphere. After production, the final product was stored in a sealed vessel preventing exposure to humidity and UV light until deposition in the AV. 

\subsubsection{Argon Purification System Components}
The argon purification system components can contribute to the backgrounds in the detector by mixing radon, deposited on or emanated from component surfaces, or other contaminants into the LAr target during filling or recirculation. The sensitive surface area of the purification loop between the radon trap and the detector is approximately 0.6~m$^2$, which includes approximately 50~welds.

To mitigate potential contamination in the process tubing, argon-wetted components of the process system were constructed with electropolished stainless steel and, where tooling was necessary, controlled TIG welding was performed with gamma-assayed and certified ceriated welding stock. Physical cleaning of stainless steel surfaces to remove dust and surface contamination, ultrasonic cleaning cycles with Alconox precision cleaner and ultra-pure water, and a chemical surface-layer contamination removal from the stainless steel and weld material with citric acid passivation were performed, followed by a final ultra-pure water rinse. After final assembly underground, the purification loop was again passivated to remove contaminants plated-out onto the surface from air exposure, and sealed until argon was introduced into the system for purification.
\section{Cryogenic System}
\label{ssec:cryo}
The cryogenic system consists of a LN$_{2}$ cooling system and a LAr purification loop. An electropolished stainless steel cooling coil, shown in Figure~\ref{fig:design}, is filled with LN$_2$ to provide the necessary cooling power to condense and maintain the argon in the detector in a liquid state between a temperature of 84--87~K and a pressure of 13--15~psia. 

The LN$_2$ is gravity-fed to the cooling coil inlet through vacuum-jacketed piping from a 3750-L storage dewar located above the detector in the Cube Hall staging area, shown in Figure~\ref{fig:InfrastructureFromModel}. Boil-off nitrogen gas is returned to the dewar where it is re-condensed by three 1-kW Stirling Cryogenics SPC-1 cryocoolers~\cite{ref:cryocooler}. During operation, two cryocoolers are operated continuously, with the third available for backup or individual shutdown during routine maintenance.

The injection of lab-temperature argon from the purification system into the AV during the initial cool-down of the AV created a high heat load on the cooling system in addition to the load due to natural boil-off from the LAr in the AV. The use of phase separators aided in preventing counter-current gas flow and vapor lock throughout the cooling system and helped maintain a controlled and stable cooling rate. 

Temperature and pressure readings are taken throughout the cryogenic delivery system, and logged using an Emerson DeltaV slow controls system~\cite{deltaV}. Valve automation, emergency shut-down, and isolation of critical components can additionally be performed remotely through this DeltaV system. 

\paragraph*{Detector Cooling Coil}
The cooling coil helix is 84 inches long and made of 0.75-inch outer diameter stock-size electropolished stainless steel tubing in a 5.50-inch center diameter helix with 35~turns at a 2.25-inch pitch. It is designed to provide up to 1000 W of cooling, even with fully submersed in LAr. The LN$_{2}$ is delivered to the bottom of the coil through a straight vacuum-jacketed supply line in the center of the helix to prevent boiling gas from flowing up the inlet line, counter-current to the downward liquid flow. The bottom of the straight LN$_{2}$ supply line curls upwards transitioning into the return helix, creating a forced convective two-phase flow heat transfer.

At the bottom of the cooling coil, flow guides were designed to promote convective LAr flow and block photons from scintillation events generated in the neck region from entering the inner AV region. As the cooling coil was designed to operate fully submerged in LAr, a detailed Computational Fluid Dynamics (CFD) analysis was performed to optimize the geometry of the flow guide, shown in Figure~\ref{fig:flowguides}.

In the final running configuration of DEAP-3600, due to a seal failure in the neck described in Section~\ref{sealFailure}, a standing column of LAr is not maintained in the neck. Instead, the LAr volume is maintained in the AV with an argon gas volume in contact with the cooling coil.  

\begin{figure}[h!]
\begin{center}
	\includegraphics[width=4in]{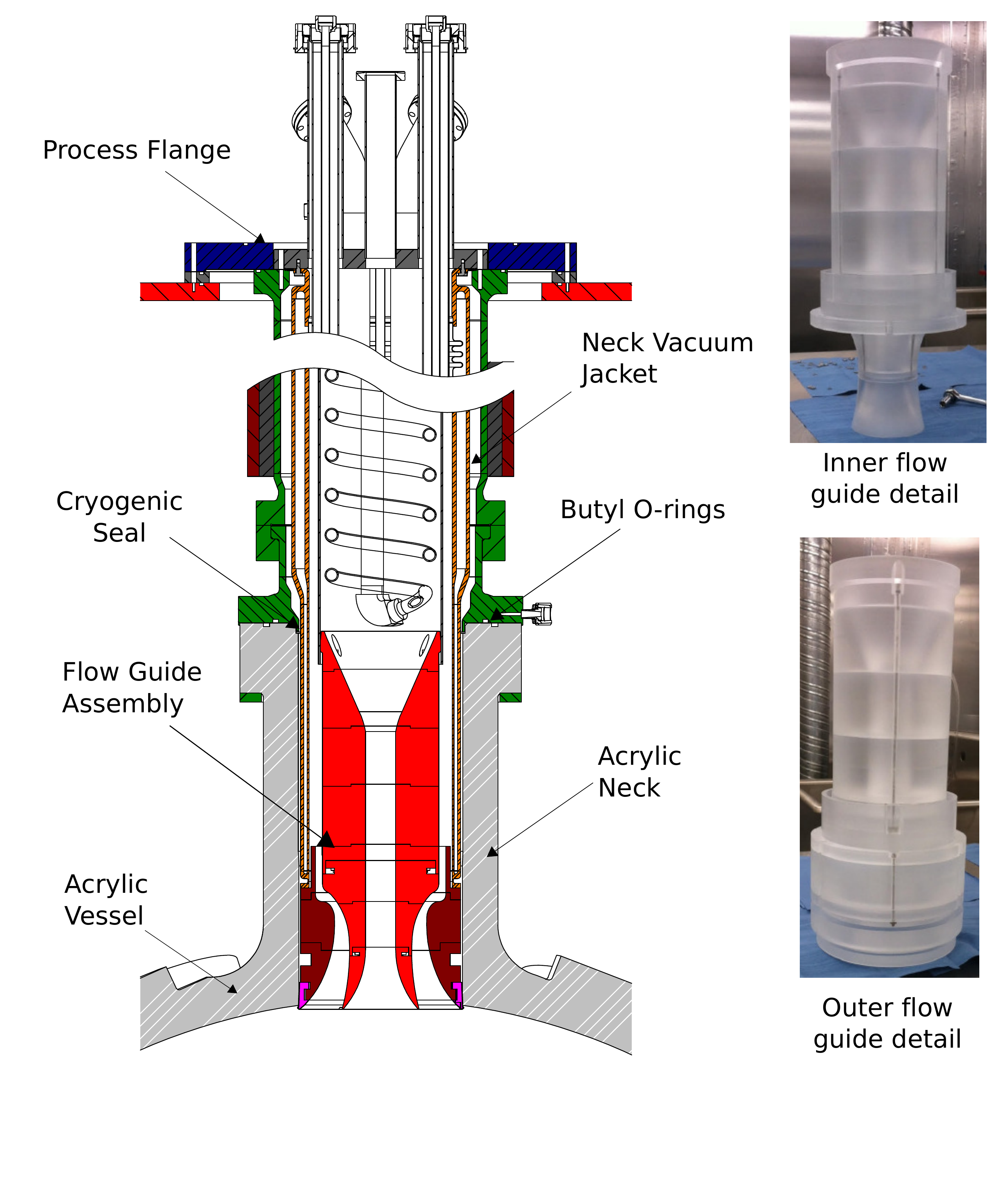}
	\caption{Left: A schematic of the internal neck. The vapour space outside the vacuum-jacketed neck (orange) and inside the outer steel neck (green) began to fill with LAr during the first AV fill, causing the failure of the butyl O-ring seals between the acrylic and steel neck interface. Right: Concentric inner and outer acrylic flow guides in AV neck designed to guide the convective liquid flow pattern. The flow guides were assembled from a stack of machined and sanded acrylic discs. A piston ring (pink) at the bottom of the flow guides covers the gap between the outer flow guide and inner acrylic neck.}
	\label{fig:flowguides}
\end{center}
\end{figure}

\subsection{Purification System}
The design goal of the purification systems is to purify the argon target to sub-ppb levels of electronegative impurities (CH$_4$, CO, CO$_2$, H$_2$, H$_2$O, N$_2$ and O$_2$) and to reduce the radon activity to as low as possible, nearing 5~$\upmu$Bq. All argon-wetted components downstream of the cryogenic radon trap were constructed from electropolished stainless steel or acrylic. For active components, such as transfer pumps and purification components, certified use of non-thoriated welding rods was demanded from the manufacturer. 

\begin{figure}[h!]
\centering
        \includegraphics[trim={0 0 0 3cm}, clip,width=4.5in]{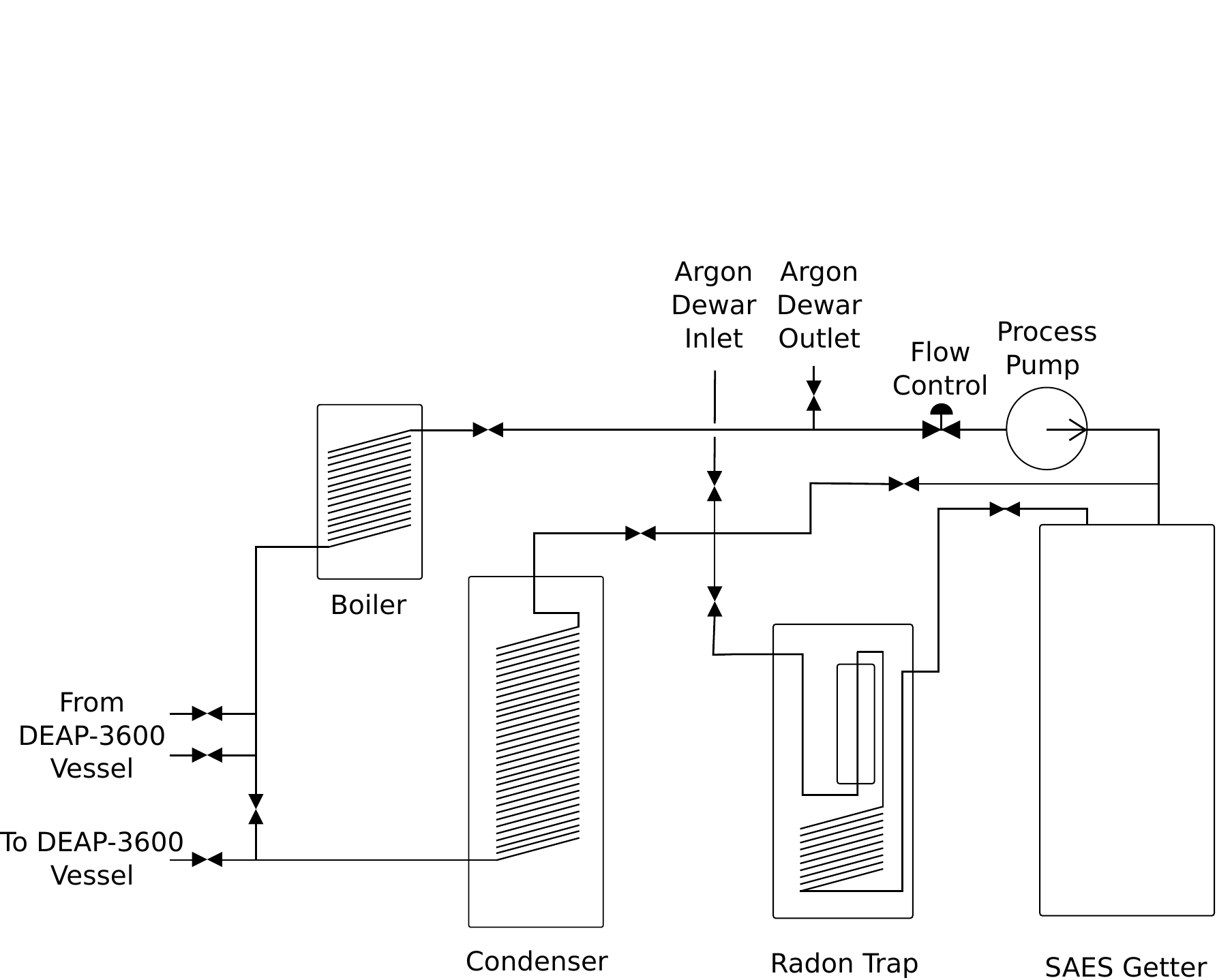}
\caption{Flow diagram of the DEAP-3600 purification system. Gas is injected into the loop ahead of a flow controller and process pump which circulates the argon through a SAES getter, radon trap, and condenser before entering the DEAP-3600 vessel. Argon returns from the vessel through a boiler unit to complete the loop.}
\label{fig:processSystems}
\end{figure}

The main purification system components, shown in a block diagram in Figure~\ref{fig:processSystems}, consist of the process pump, SAES getter, and custom-built radon trap, condenser column, and boiler. The system was designed for a maximum nominal flow rate of 4.9~g/s of argon and accepts argon gas at 300~K from a bulk liquid storage tank. 

Gas is injected into a KNF Neuberger~150.1.2.12 double diaphragm process pump, which maintains a forward pressure of 30~psig at the top of the system. The double diaphragm model safely imposes an extra barrier between the lab air and process gas in the event of pump failure, and the pressure between the two diaphragms is monitored. 

Chemical purification is performed by a SAES Mega-Torr PS5-MGT15 hot metal getter custom fabricated for DEAP to avoid internal components with thoriated welds. It is specified to accept 99.999\% high-purity argon at a maximum flow rate of 7.4~g/s. The getter contains a safety interlock system which prevents over-heating and ignition of the getter material in the attempt to purify gas that is too high in electronegative impurities.

Radon and radioactive impurities are removed by absorption in a custom-built charcoal trap, designed to take gas at 300~K from the getter, pre-cool it to 100~K, and pass it through a charcoal column. For optimal performance, the trap should be as cold as possible while maintaining argon in the gas phase. The trap is placed between  the active purification system components and the detector to minimize emanated radon from the system itself from mixing into the argon volume. The inlet is surrounded by a copper block, partially immersed in LN$_2$, with tunable cartridge heaters, capable of 600~W of heating, to prevent argon gas from freezing. The charcoal cartridge is a 12-inch cylinder with a 3-inch diameter filled with 610~g of Saratech charcoal, selected to have very low radon emanation. The charcoal is contained by a barrier of stainless steel wool, retaining steel mesh, and 50-$\upmu$m VCR filter gaskets on both the top and bottom to prevent particulates from escaping. The cartridge is surrounded by a bake-out heater. All components of the radon trap are contained within an 8-inch-diameter cylinder, wrapped in multi-layer insulating foil, and housed within a 10~inch vacuum space. 

The custom-built condenser column is built to liquify gas from the radon trap before delivering it into the detector. The condenser comprises a stainless steel coil formed from a 39-ft long, 0.5-inch-outer diameter stainless steel tube, suspended inside an 8-inch-diameter cylinder and immersed in LN$_2$. Either liquid or gaseous argon can be delivered to the AV; for gaseous argon it is liquified on the neck cooling coil before dripping into the AV. 

Purified argon gas is directed into the AV via the inlet on the main process flange. The system was designed so that liquid could be extracted and then vaporized in a boiler before being returned to the purification system. In the current configuration, with gas in the detector neck, gas is returned to the boiler inlet. An auxiliary gas return, running in parallel to the main outlet also connects the process flange to the purification loop before the boiler unit. To avoid direct exposure of the heating elements to the argon, a 1.5-kW heater is coiled around the stainless steel flow line returning from the AV to vaporize the LAr. Gas exiting the boiler is delivered back through the flow control valve into the KNF pump at the the top of the loop.

If any of the relief valves throughout the system opens and fails to close fully, it is possible to back-stream lab air into the purification system. To mitigate this potential contamination, a double check valve assembly is used on all pressure relief assemblies, shown in Figure~\ref{fig:reliefAssembly}. A 3~psig check valve is in contact with the argon gas. Behind this inner check valve is a small enclosed volume with a valve to purge with argon gas, pressure gauge to indicate a pressure relief, and a variable pressure outer relief valve to provide a redundant seal. 

\begin{figure}[h!]
\begin{center}
	\includegraphics[width=4.5in]{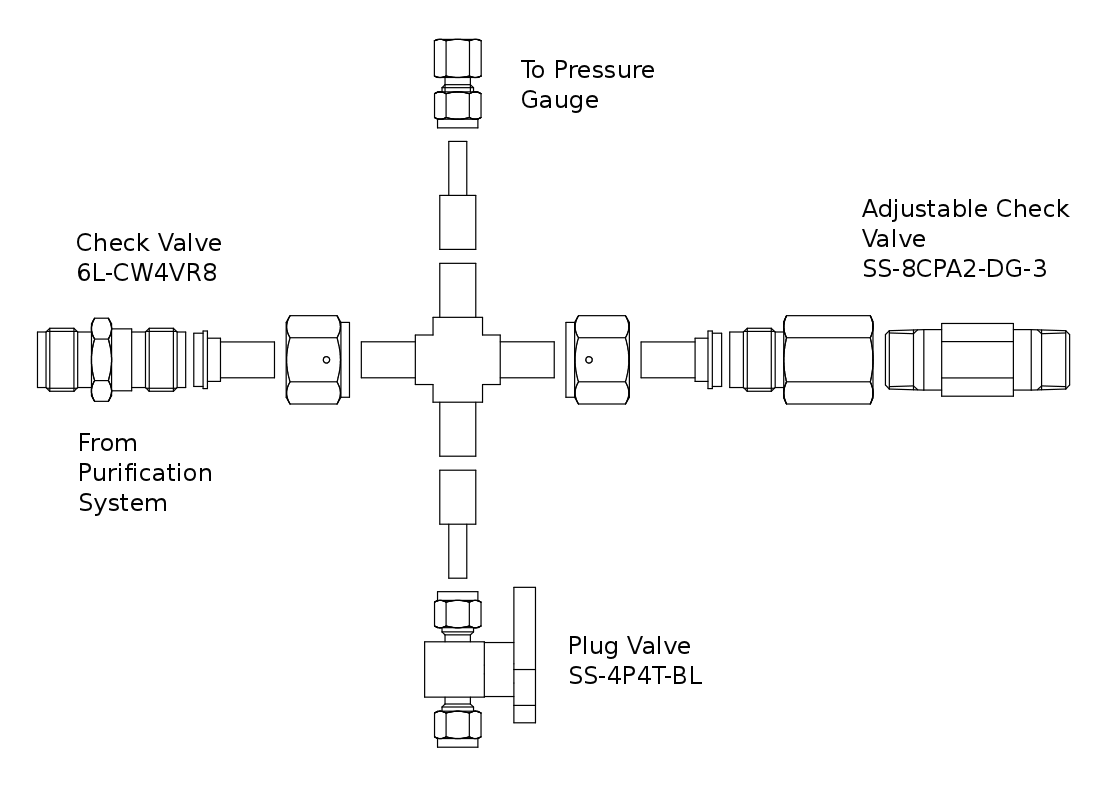}
	\caption{Double check valve relief assembly used throughout the purification system. A check valve in direct contact with the purification system leads to small volume back-filled with clean argon gas, a pressure gauge, purge valve, and variable pressure relief valve.}
	\label{fig:reliefAssembly}
\end{center}
\end{figure}

The process systems may be operated in 3 main configurations: filling, recirculation, and storage recirculation. After the initial purification of argon when filling the detector, after one year of running, re-purification of the argon has not be necessary. 

\textbf{Filling} ---  During filling, gas is constantly drawn from storage by the purification system, purified, and injected into the detector. The gas can be liquified before delivery into the AV or can be injected as gas and liquified by the cooling coil. The injection rate can be varied between 0--4.9~g/s depending upon the cooling rate of the AV and the available cooling capacity. 

\textbf{Recirculation} --- After filling, or during pauses, the system may be operated to recirculate boil-off gas around the process loop, re-purifying and liquifying in a steady state. If the detector is full, liquid may be extracted directly, boiled, and recirculated around the loop.

\textbf{Storage recirculation} --- The detector may be bypassed and cold purified gas extracted after the radon trap and transferred back to the 3750-L storage dewar. This is the default mode of the system, allowing for stabilization of the purification system during start-up and pre-purification of the gas space in the storage dewar. 

\subsection{Neck Seal Incident}\label{sealFailure}
A vapor space exists between the outer vacuum-jacketed neck (shown in orange in Figure~\ref{fig:flowguides}) and the inner surface of the stainless steel neck (green). There is a small fit-tolerance between the vacuum-jacketed neck and the inner surface of the acrylic neck (grey). The piston ring (described and shown in pink in Figure~\ref{fig:flowguides}) does not produce a true seal at the bottom of the neck to prevent LAr from filling into the vapor space.

During the initial filling of the detector, the LAr level rose through the acrylic neck and flow guides. A leak at the connection between the process flange (blue in Figure~\ref{fig:flowguides}) and inner vacuum-jacketed neck prevented the hydrostatic head pressure needed to keep LAr from filling the outer space between the vacuum-jacketed neck and acrylic neck. The rising LAr filled equally the inner space contained by the flow-guide assembly and the space between the outer vacuum-jacket and inner acrylic neck, coming in direct contact with the acrylic. 

The acrylic to steel neck interface is sealed with 2 butyl O-rings and an additional cryogenic seal, shown in Figure~\ref{fig:flowguides}, designed to contract and seal when slowly cooled. The rapid temperature drop due to direct exposure to LAr in the neck acrylic lead to failure of the butyl seals, allowing a pathway for clean, radon-scrubbed boil-off nitrogen that was purging the steel shell volume to leak into the AV volume. On 17~August 2016, a contamination of approximately 100~ppm N$_2$ leaked into the LAr causing a sharp decrease in the observed long time constant to argon scintillation and a spike in the AV pressure. This required complete venting and boiling of the approximately 3600~kg LAr in the AV. 

After the AV was emptied of all cryogen and the cause of the seal failure identified, clean Ar gas was injected and liquified in the AV. A reduced fill level was chosen to minimize the possibility of LAr reaching the butyl seals again. A final fill level corresponding to approximately 3300~kg LAr was chosen and has been stably maintained since completing the second fill in November 2016. 
\section{Inner Detector Construction}\label{ssec:ID}
The inner detector consists of all elements between the LAr and the steel shell, as summarized in Figure~\ref{fig:innerDetPic}. The construction of these components is described in this section, with the exception of the PMTs (see Section~\ref{ssec:PMT}).

\begin{figure}[htb]
\centering
     \includegraphics[width=\textwidth]{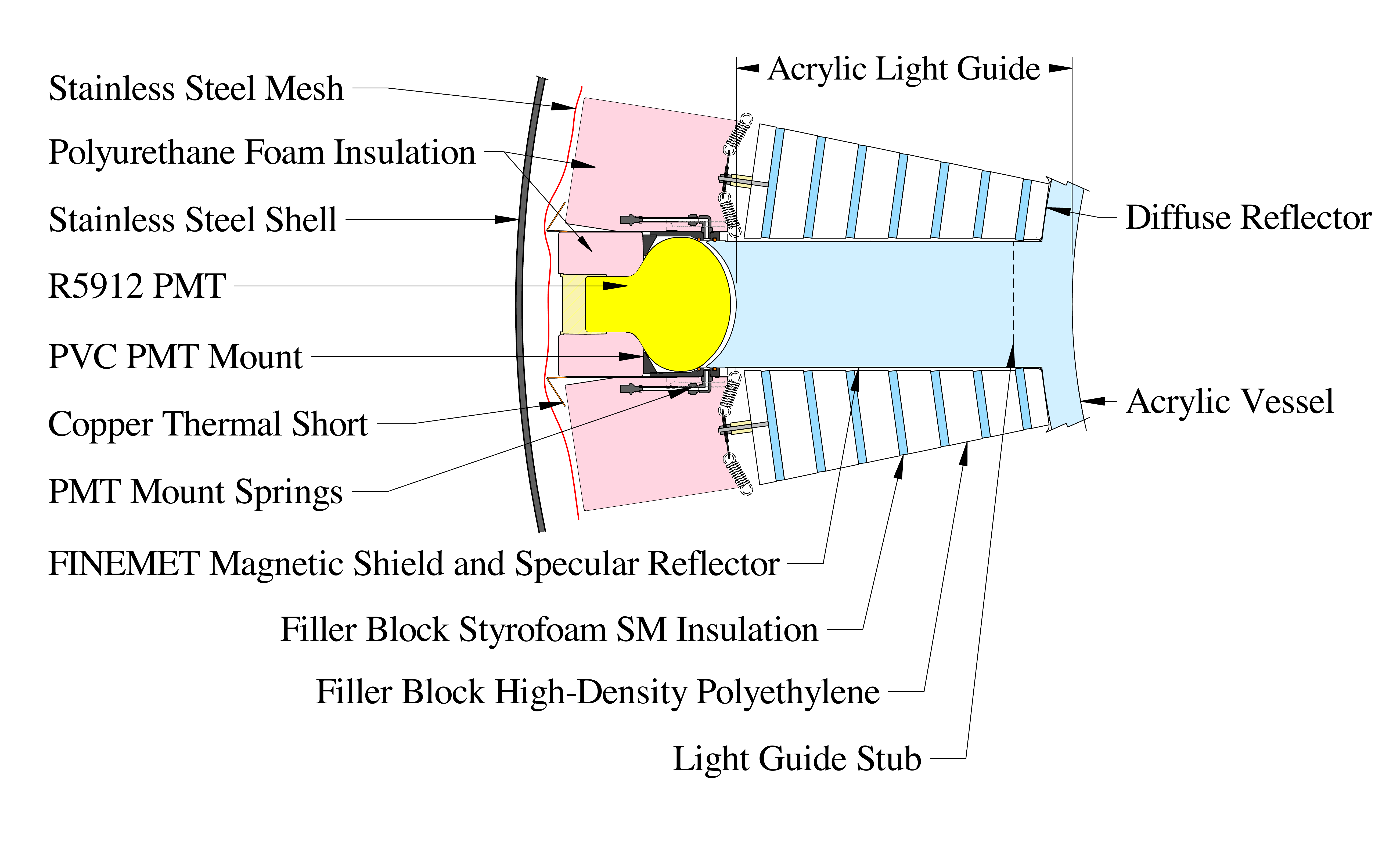}
     \caption{Cross section of the inner detector components from the acrylic vessel to the stainless steel pressure vessel.}
     \label{fig:innerDetPic}
\end{figure}

\paragraph*{Acrylic Vessel}
Cast acrylic is mechanically strong enough that few metal supports are required to hold the weight of the AV, LAr, LGs, filler blocks, and PMTs. The AV was built in three pieces (the neck, collar, and the truncated sphere) due to the limited envelope of the mine shaft leading underground. The average vessel radius is estimated at 846~mm when cold and 7~mm larger at room temperature. The neck provides mechanical support of the AV from above and access for the purification system cooling coil. The inner-neck diameter is 255~mm.

Each of these three acrylic pieces underwent a different construction:
\begin{itemize}
	\item The neck raw material was made from acrylic sheets bonded together to form a rough cuboid.
	\item The collar was made of a single, thermoformed panel of acrylic.
	\item The truncated sphere consisted of 5 spherical slices and one polar cap, thermoformed and then bonded together by RPT.
\end{itemize}

The acrylic sphere was machined to the final spherical shape, shown in Figure~\ref{fig:Machining}, on a 5-axis computer numerical controlled (CNC) mill at the University of Alberta. LG stubs were then machined to receive the LGs. Particular care was given to avoid unnecessary exposure of the acrylic to any material that might generate surface stress, such as methanol and MMA. The machining was performed using the guidelines from Stachiw~\cite{Stachiw2003}.

\begin{figure}[h!]
        \centering
        \includegraphics[width=1.0\textwidth]{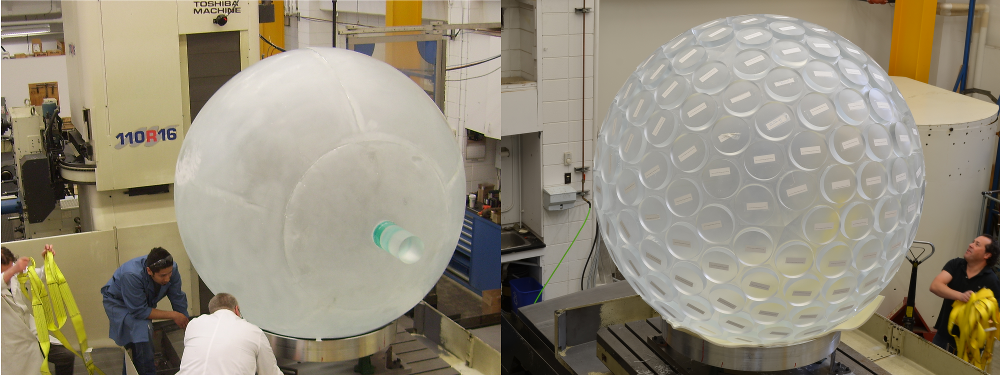}
	\caption{AV as delivered by Reynolds Polymer Technologies (left) and after machining (right) at the University of Alberta.}
	\label{fig:Machining}
\end{figure}

The AV sphere, collar, and neck were then shipped to the SNOLAB underground facility and bonded together by RPT. Due to the large tolerances produced in the bonding process, additional post-machining underground was required.

\paragraph*{Flow Guides}
The acrylic used to construct the AV was also used for the flow guide assembly, shown in Figure~\ref{fig:flowguides}. These were milled at the University of Alberta in a controlled room with a radon level of 0.3~mBq/m$^3$. Once machined to shape, approximately 20~$\upmu$m was removed by hand-sanding in a nitrogen-purged glovebox at Queen's University with an oxygen content below 20~ppm.

\paragraph*{Light Guides and PMT Mount Assembly}
Each of the 255 light guides is 45~cm long and 19~cm in diameter, giving an AV surface coverage of 76\%. The outer LG face is concave to match the geometry of the PMT face. The annealing, machining on CNC lathe, and polishing of the LGs were completed at TRIUMF in British Columbia, Canada.

The LGs were bonded on the AV underground using a tripod set in position with an alignment cylinder suctioned to the target stub, and attached to the AV by clamping to neighboring stubs or LGs, shown in Figure~\ref{fig:Bonding}. The LG to be bonded was constrained in the tripod to only move perpendicularly to the stub face with a compression spring.

\begin{figure}[htbp]
        \centering
        \includegraphics[width=0.6\textwidth]{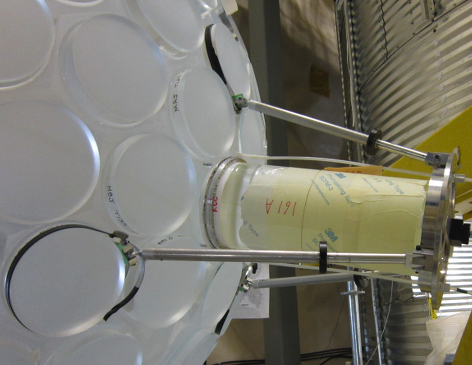}
	\caption{A light guide (right) ready to be bonded on the AV (left).}
	\label{fig:Bonding}
\end{figure}

The bond was created by constructing a form-fit polyethylene dam with a radial bulge around the outer perimeter of the LG-stub gap interface. This bulge allowed for an excess reservoir of acrylic monomer to prevent concavity in the bond during polymerization, and was machined off after a post-bonding anneal. The dam was sealed against the stub with O-ring clamps. A computer-controlled fill system was used to ensure consistent bonds. 

Although approximately 80\% of photons emitted in the LAr are trapped in the LGs by total internal reflection, an additional specular reflector, loosely wrapped around the LGs, is used to increase light collection and provide optical isolation between LGs. Aluminized mylar was selected for its high reflectance and lack of alpha-induced scintillation. A 50-$\upmu$m mylar foil was sputter-coated by Astral Technology Unlimited (Northfield, MN, USA) with 100~nm of aluminum, using a 99.999\% high purity custom-made sputtering target by Laurand Associates (Great Neck, NY, USA) to satisfy the radiopurity requirements.

Areas on the spherical AV shell between LGs were covered in a diffuse, 98\% reflective white Tyvek base layer followed by a layer of black Tyvek and a layer of closed-cell polyethylene foam backing. These layers help to maximize the light collection and keep stray photons originating outside the active volume from leaking into the LAr volume.

At the end of each LG, a PMT mount assembly, detailed in an exploded view in Figure~\ref{fig:pmtAssembly}, supports and maintains optical coupling between the PMT and LG. A cylindrical PVC barrel seals to the end of the LG with an O-ring and a lightly-compressed backing yoke. The LG--PMT interface volume is filled with silicone oil (Sigma Aldrich $\#$378399) serving as an optical coupling, which was found to be a good index match between acrylic and glass, and has a favorable coefficient of thermal expansion and viscosity. The assembly is wrapped in a sleeve of FINEMET~\cite{ref:finemet} magnetic shielding (Section~\ref{subsec:magcomp}) and an outer sleeve made of copper acting as a thermal short to passively warm and prevent thermal gradients across each PMT. 

\begin{figure}[h!]
\centering
	\includegraphics[width=4.7in]{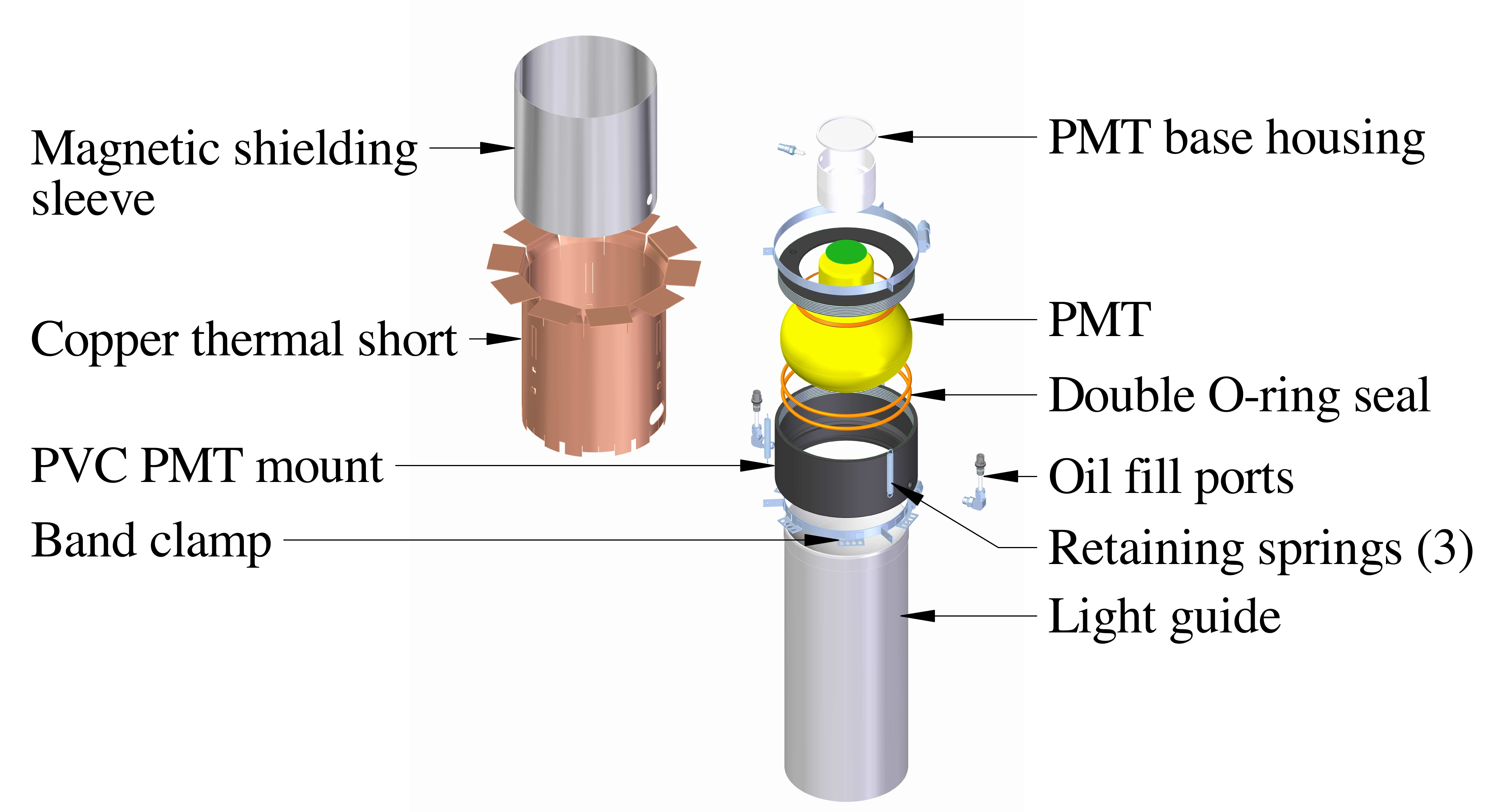}
	\caption{Exploded view of PMT mount assembly and light guide.}
	\label{fig:pmtAssembly}
\end{figure}

\paragraph*{Filler Blocks} \label{subs:fillerblocks}
The volume between LGs is filled with 486~filler block stacks fabricated from alternating layers of high-density polyethylene and Styrofoam~\cite{ref:styrofoam}. The polyethylene and Styrofoam combination provides superior thermal insulation and equivalent neutron shielding compared to acrylic alone. Retaining springs at the warm end of the LG push the filler blocks against the AV and keep the blocks centered between LGs when the acrylic contracts during cool-down. A 5~mm gap between the LGs and filler blocks ensure adequate spacing for thermal expansion without generating stress. 

\paragraph*{Temperature Sensors, Foam Insulation, and Steel Mesh} \label{subsec:FoamInsulation}
The inner detector is instrumented with an array of 85~PT-100 resistive temperature detectors. For each selected filler block, a sensor was bonded to the bottom, middle and top block layers, for temperature measurement at a distance of 0.9~m, 1.1~m, and 1.3~m, respectively, from the center of the AV. A fourth sensor was bonded to the copper thermal short on a LG adjacent to the filler block.

The copper thermal shorts are surrounded by pieces of open cell polyurethane foam for additional thermal insulation. A stainless steel mesh is fastened around the entire detector to contain shattered components in the event of an AV structural failure that could block the vapor relief path along the steel neck. The volume behind the mesh and inside the steel shell is maintained as a vapor space, and is continuously purged with radon-scrubbed boil-off nitrogen gas.

Figure~\ref{fig:InnerDet1} shows the stages of installation during construction of the PMTs, filler blocks, and foam insulation on the AV.

\begin{figure}[htb]
\begin{centering}
     \includegraphics[width=0.9\textwidth]{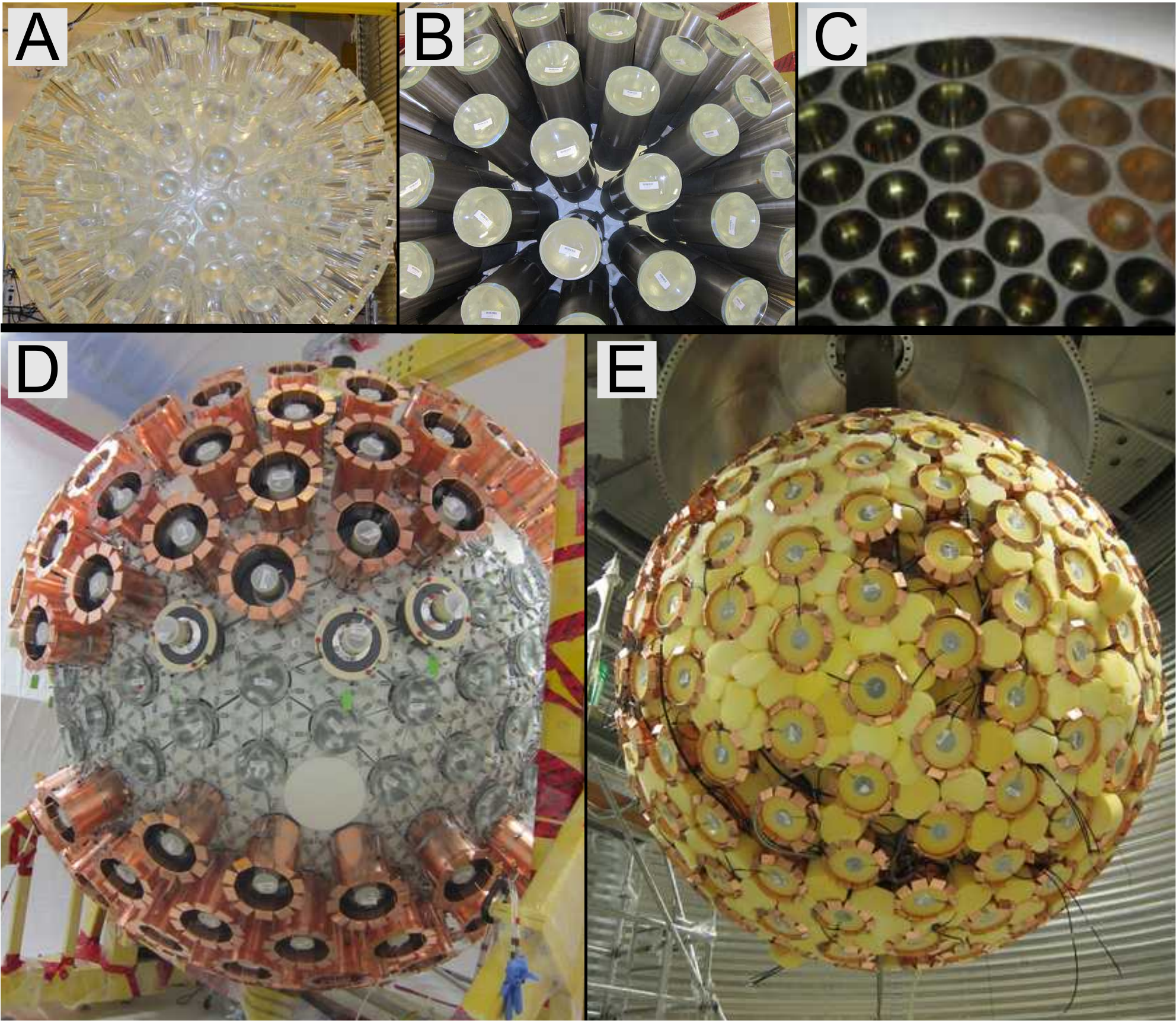}
     \caption{A)~The acrylic vessel after bonding on the light guides. B)~Reflectors and magnetic shielding installed around light guides. C)~View from inside the detector with the white Tyvek and most PMTs installed on light guides. D)~Detector with filler blocks installed and during PMT installation. E)~Detector with all PMTs installed and during backing foam installation.}
     \label{fig:InnerDet1}
\end{centering}
\end{figure}

\subsubsection{Acrylic Treatment}
\paragraph*{Annealing} \label{subs:annealing}
The AV was annealed five times during its construction underground at a temperature between 80$^{\circ}$C and 85$^{\circ}$C to relieve stress in the material after bonding and to harden bonds. These anneals occurred (i) after machining of the AV truncated sphere, shoulder, and neck; (ii) after bonding the AV truncated sphere, shoulder, and neck together; (iii) after underground machining of the neck stubs, neck and shoulder bond areas, and before LG bonding; (iv) after LG bonding and before bond-bulge removal; and (v) after all bonding was completed, a final extra anneal. A 10-ft cubical oven was constructed from foam insulation panels to perform the annealing. An external heater and blower circulated heated air in the oven. Several smaller ducts were positioned across the oven floor to distribute the reheated air. The difference in temperature across the oven was required to be less than 2${}^{\circ}$C for annealing. 

As radon diffusion increases with temperature, the level of radon in the air inside the AV was monitored and controlled to below 10~Bq/m$^3$ when annealing. During the first three anneals, heated air that was piped underground from the surface was blown into the vessel. This surface-air is lower in radon by approximately a factor of 10~with respect to the underground lab air. After bonding the neck to the AV, the inside volume was flushed with lab-grade nitrogen (fourth cycle) or argon (fifth cycle) and then kept sealed during annealing. Using the exposure history and concentration measurements at each stage of construction and annealing, the calculated build up of radon progeny from the integrated radon exposure of the AV would result in a \Nuc{Pb}{210} activity of 14~mBq/kg in the surface acrylic prior to resurfacing.

\paragraph*{Acrylic Resurfacing}\label{sec:resurfacer}
A robotic sander, the ``resurfacer" shown in Figure~\ref{fig:resurfacer}, was designed to remove up to 1~mm of the inner surface layer of the spherical AV. Two rotating sanding heads with 3M Flexible Diamond QRS Cloth Sheet 6002J M74 sanding pads were drawn across the inner surface of the AV to remove and collect the sanded acrylic under a continuous flush with ultra-pure water. When sanding, a linear actuator pushed the sanding motor outwards against the acrylic, developing a constant, nominal 12-lb normal force against the AV.

In each sanding pass, the tilt arm moved in a spiral pattern from its deployed vertical position down to the equator and back, with the opposing arm section sweeping out the region between the south pole and equator. The starting azimuthal angle was incremented by 60$^{\circ}$ before the start of the next pass to provide a more uniform coverage across the sanding surface. After sanding of the surface layer was completed, the sanding heads were retracted, and the acrylic surface was flushed with ultra-pure water to extract remaining loose material.

\begin{figure}[h!]
\centering\hspace*{+0.6in}
\includegraphics[width=3.7in]{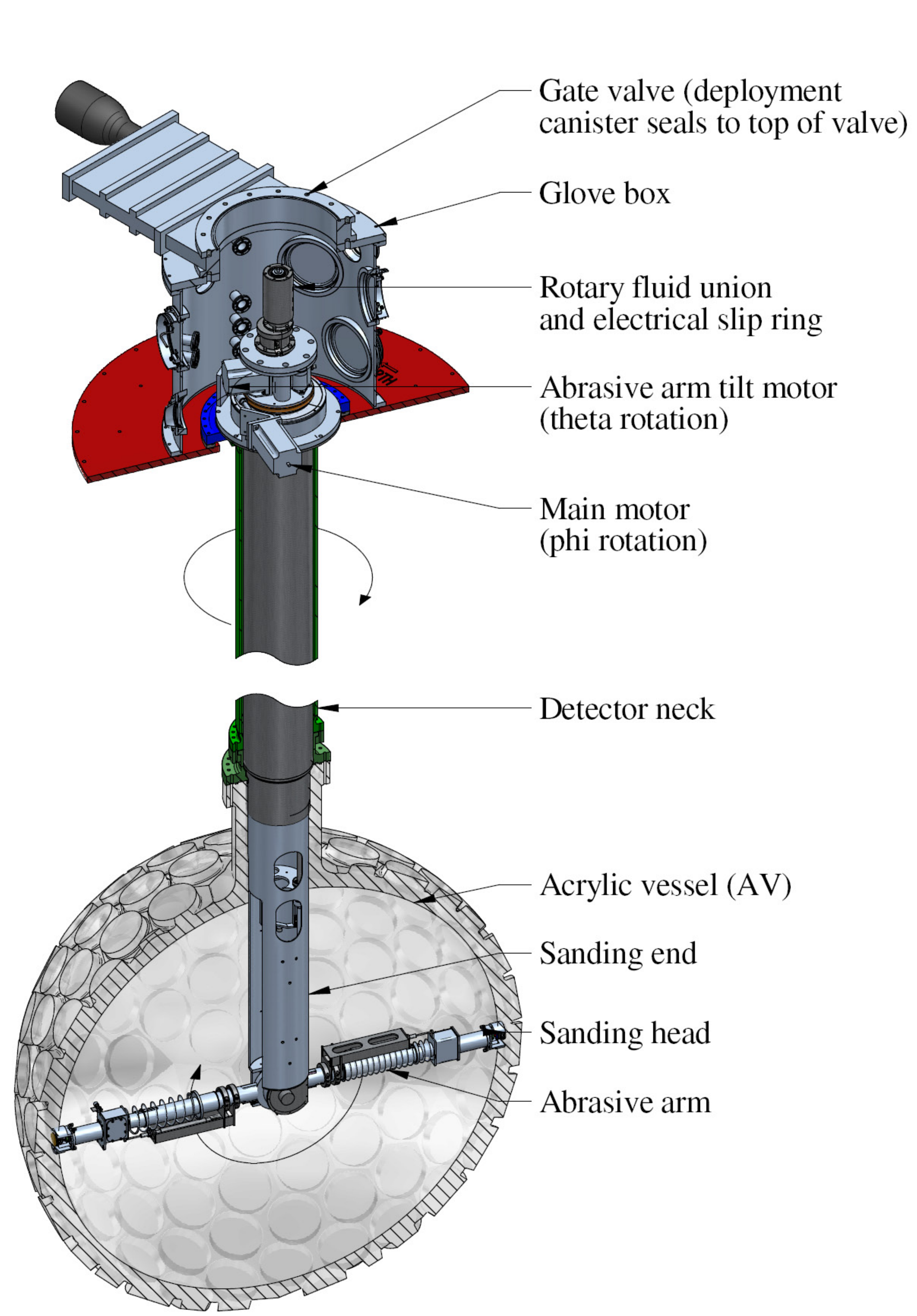}\vspace*{+0.1in}
\caption{Overview of the resurfacer device when deployed in the acrylic vessel.}
\label{fig:resurfacer}
\end{figure}

The resurfacer was deployed through the AV neck and operated with the detector volume hermetically sealed from the lab and continuously purged with radon-scrubbed boil-off nitrogen gas~\cite{wojcik2005low}. After sanding, the resurfacer was extracted through the glovebox and into an auxiliary canister which was sealed and purged with radon-scrubbed boil-off nitrogen, avoiding any exposure to laboratory air. 

The integrated sanding time over the inner AV surface was approximately 198~hours. Based on this sanding time and the measured acrylic removal efficiency, an estimated $500\pm 50~\upmu$m of acrylic was removed from the inner surface of the AV. This thickness is sufficient to reduce the \Nuc{Pb}{210} surface backgrounds down to near the upper assay limit of $2.2 \times 10^{-19}$g/g (0.62~mBq/kg), as shown in Figure~\ref{fig:acrylicRemoval}.

\begin{figure}[h!]
\begin{center}
\includegraphics[width=5in]{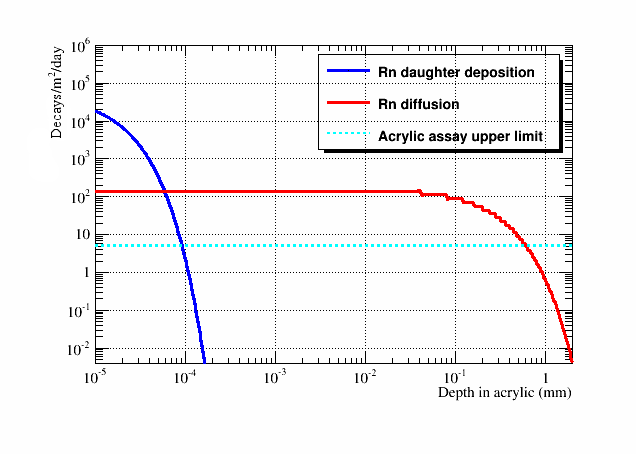}
\caption{Calculated \Nuc{Pb}{210} alpha activity in the AV before resurfacing. Activity after radon-laden air exposure (9 months on surface, 6 months in mine air and 1 month in radon reduced air) is shown in blue. Activity due to radon diffusion into acrylic is shown in red. The cyan line is from acrylic assay upper limit of $2.2 \times 10^{-19}$g/g in the AV. The $500~\pm~ 50~\upmu$m of removed acrylic reduces the activity of \Nuc{Pb}{210} down to near the assay upper limit.}
\label{fig:acrylicRemoval}
\end{center}
\end{figure}

\paragraph*{Wavelength Shifter Deposition}\label{sec:TPB Deposition}
An 11-cm-diameter spherical evaporation source was constructed from 316 stainless steel, shown in Figure~\ref{fig:TPBsource}, for the application of the TPB wavelength shifter to the inner surface of the AV. Details on the prototyping and testing of the source can be found in~\cite{Pollmann2012} and~\cite{BroermanThesis}. An inner copper crucible, which holds the TPB powder, was radiatively heated by a flexible Watlow 125CH93A1X coil heater wound on the outside of the evaporation source. As the source was heated, the TPB molecules scattered inside the source before exiting through one of the twenty 14-mm-diameter holes, creating a uniform outgoing flux. 

\begin{figure}[!ht]
\begin{center}
	\includegraphics[width=1.75in]{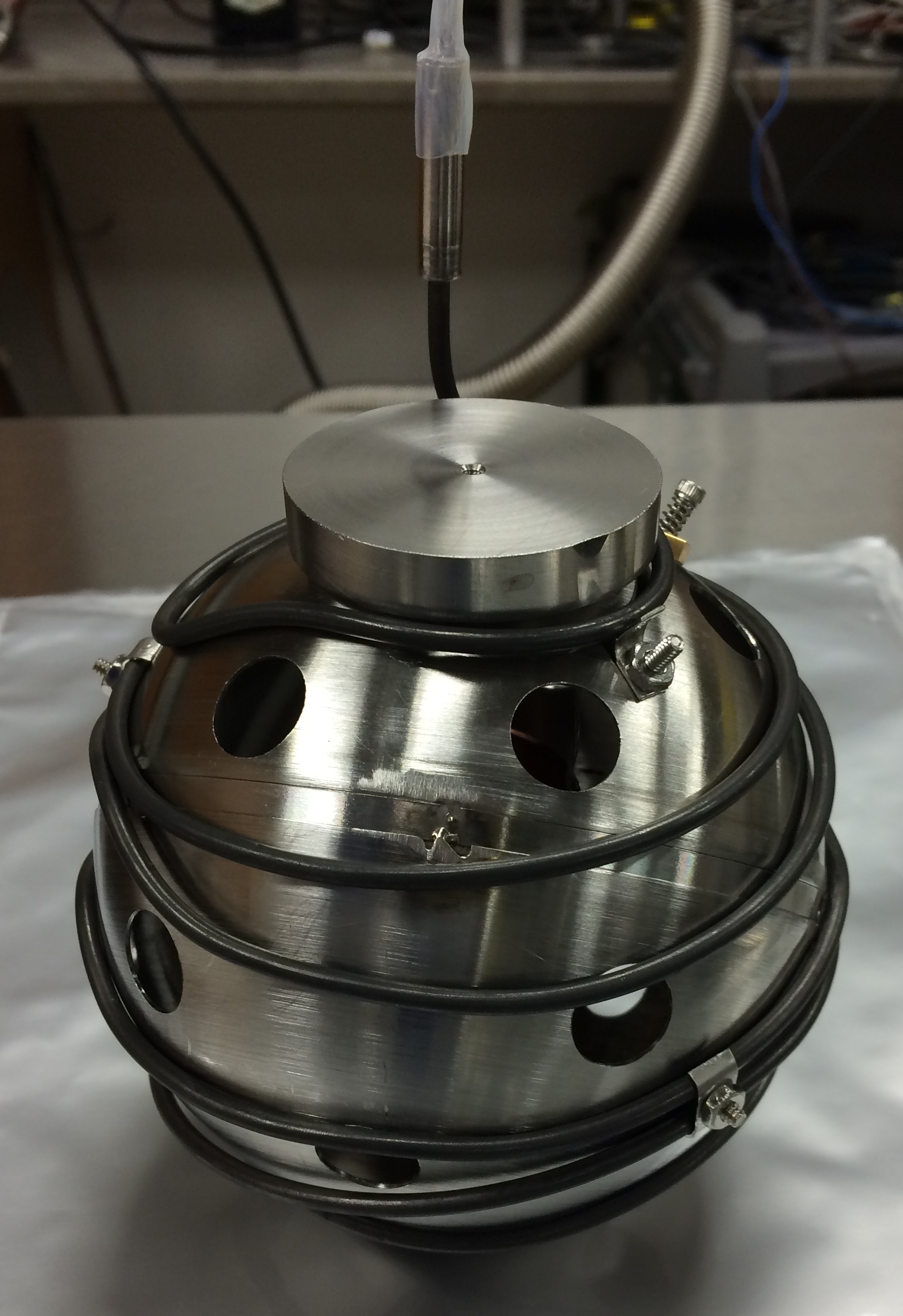}
\caption{TPB evaporation source. A heating wire wraps the outside of the sphere, radiatively heating the copper crucible, which is contained within the sphere, evaporating the TPB powder.}
\label{fig:TPBsource}
\end{center}
\end{figure}

After the resurfacing of the AV, the empty TPB source was deployed to the center of the AV for a vacuum bake of the acrylic. The inner surface of the AV was brought to 50${}^{\circ}$C to outgas absorbed water and reach the approximately $10^{-6}$~mbar vacuum needed for the deposition. Two evaporations of 29.4~$\pm$~0.2~g combined total mass were performed on the AV, which yielded a uniform coating thickness of 3.00~$\pm$~0.02~$\upmu$m. A detailed description of the TPB deposition can be found in~\cite{broerman2017application}.
\section{Light Detection Systems}\label{sec:LightDetection}

\subsection{Photomultiplier Tubes}\label{ssec:PMT}
Hamamatsu R5912 8-inch-diameter HQE PMTs~\cite{ref:r5912} were selected for DEAP-3600 for their high photon detection efficiency (nominal 32\%), low dark noise rates, and good timing characteristics. Details on the PMT characterization can be found in~\cite{pmtpaper}, while ensemble characterization versus PMT number (PMTID), which is closely coupled to the vertical location of the PMT, for all 255~LAr PMTs is presented herein.

The PMTs operate at bias voltages between 1500~V and 1900~V, described in Section~\ref{sec:electronics}.  In-situ measurement of the mean single photoelectron (SPE) charge for all PMTs is shown in Figure~\ref{fig:pmtchar1} and has a mean of  9.39~pC and a RMS of 0.16~pC. Two outliers are PMTs that developed faults in the base. The mean SPE charges are monitored on an ongoing basis and are related to the applied bias voltage through:
\begin{equation}\nonumber
\overline{q} = \text{A}\cdot \text{V}^\gamma \label{eq:gainvsbias}
\end{equation}
where $\overline{q}$ is the mean SPE charge, V is the bias voltage, and A a normalization parameter. The $\gamma$ parameter was measured for most PMTs\footnote{Data were not available for 36~PMTs as some were kept at their nominal voltage to verify that the LED light intensity did not vary between runs, and some DAQ channels had not yet been configured.} and is also shown in Figure~\ref{fig:pmtchar1} with a distribution mean of 6.9~with an RMS of 0.2.

\begin{figure}[h!]
\begin{center}
	\includegraphics[width=3.5in]{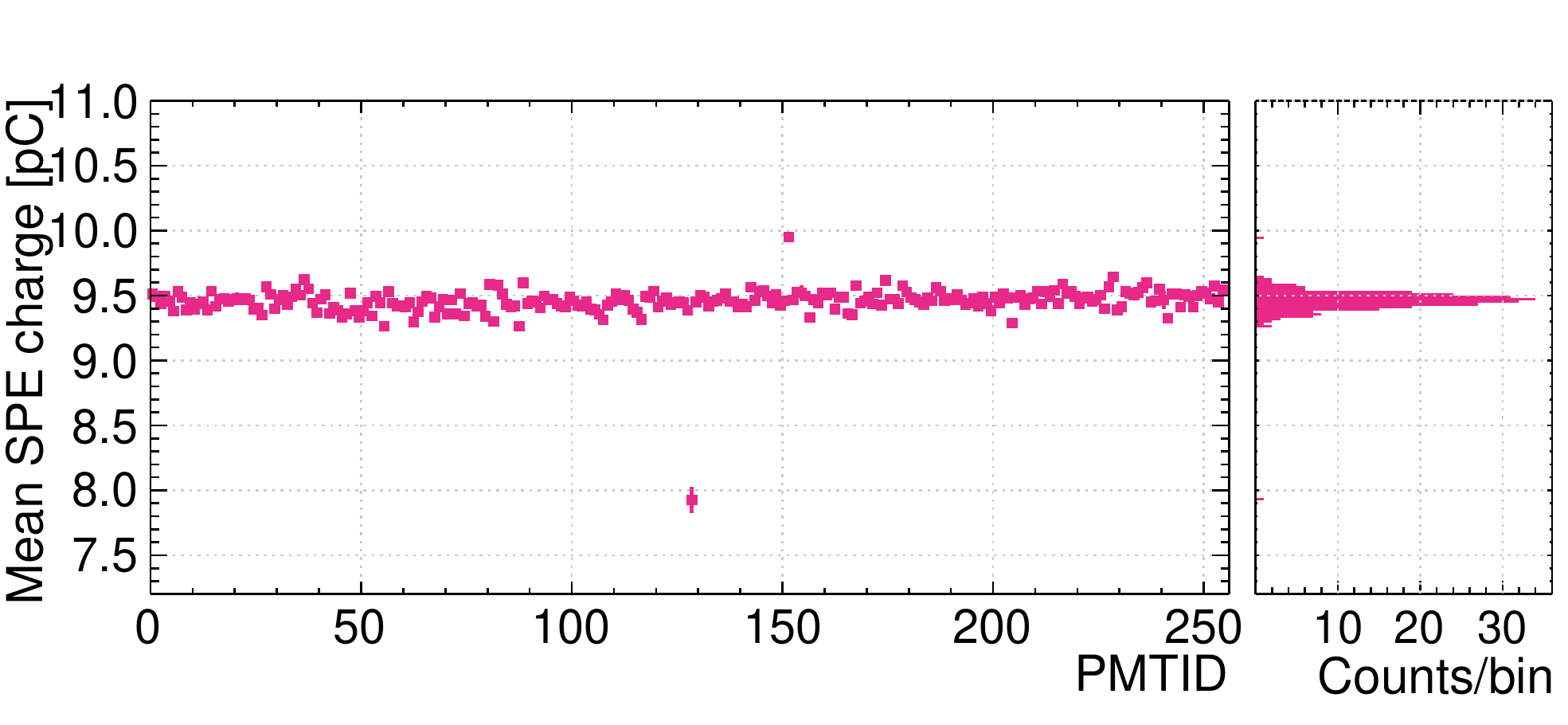}
	\includegraphics[width=3.5in]{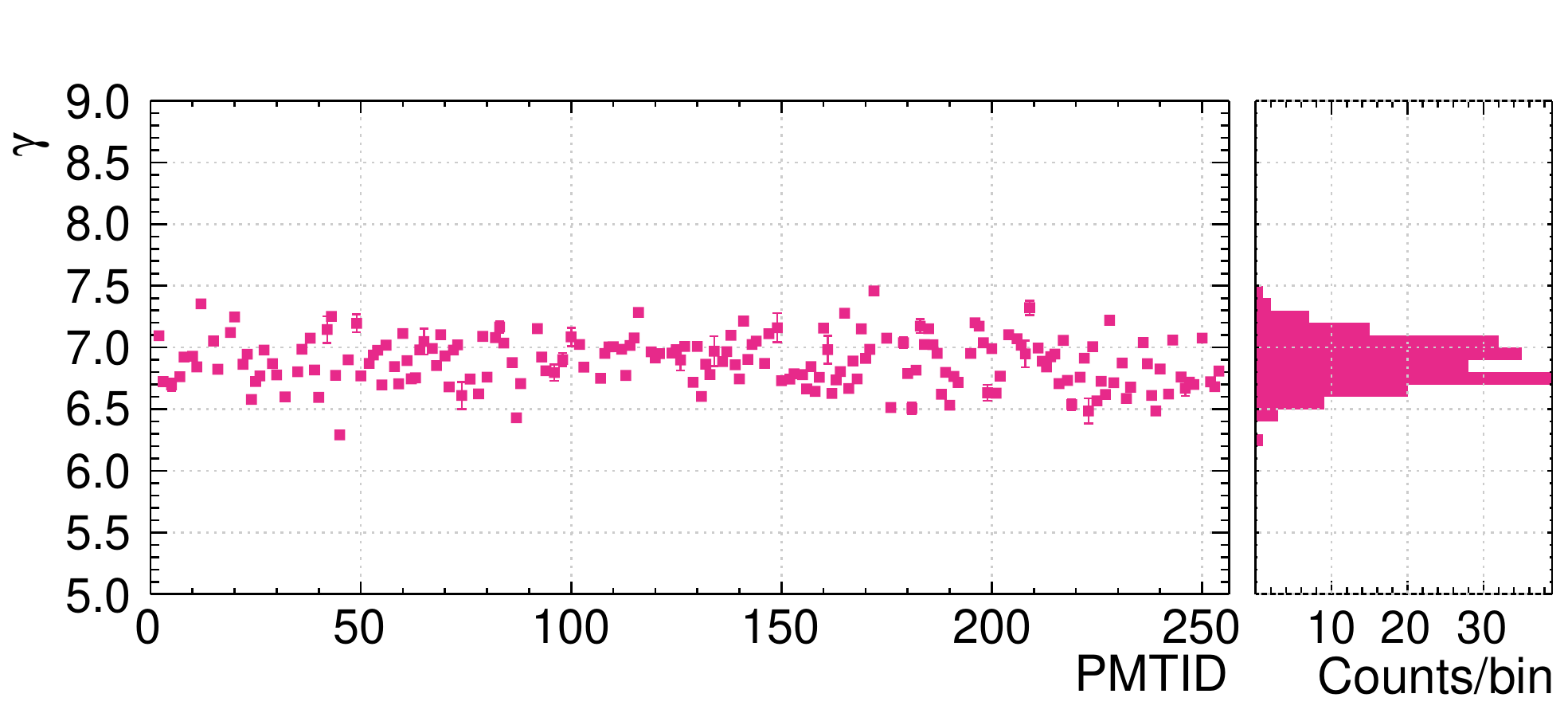}
     \caption[]{Top: Mean single photoelectron charge vs.~PMTID with a mean of 9.39~pC and an RMS of 0.16~pC. Bottom: $\gamma$ parameter vs.~PMTID, with a mean of 6.9~and RMS~of 0.2, that relates the mean SPE charge to the bias voltage is shown vs. PMTID.}
	\label{fig:pmtchar1}
\end{center}
\end{figure}	

The dark noise rates per PMT, shown in Figure~\ref{fig:pmtchar2}, are strongly temperature-dependent and are therefore shown while the detector was at room-temperature and right after cool-down when the PMTs were close to their operating temperature, with those near the top of the detector at 280~$\pm$~2~K and those near the bottom at 260~$\pm$~2~K.  The data were taken while argon gas was inside the AV. The room-temperature dark noise rate has a mean of 5.80~kHz and an RMS of 0.78~kHz. The dark noise distribution with cold argon gas has a mean of 0.24~kHz and an RMS of 0.06~kHz. When the detector is filled with LAr and the PMTs are at their operating temperature, the true dark noise rate cannot be measured due to the high rate of \Nuc{Ar}{39} events.

The full-width at half-maximum (FWHM) transit time spread, due to the variance in time from photoelectrons (PEs) liberated off the photocathode to impingement on the first dynode, was measured ex-situ using a tagged \Nuc{Sr}{90} source. The distribution, shown in Figure~\ref{fig:pmtchar2}, has a mean of 2.60~ns across all 255~PMTs with an RMS of 0.12~ns. 

\begin{figure}[h!]
\begin{center}	
	\includegraphics[width=3.5in]{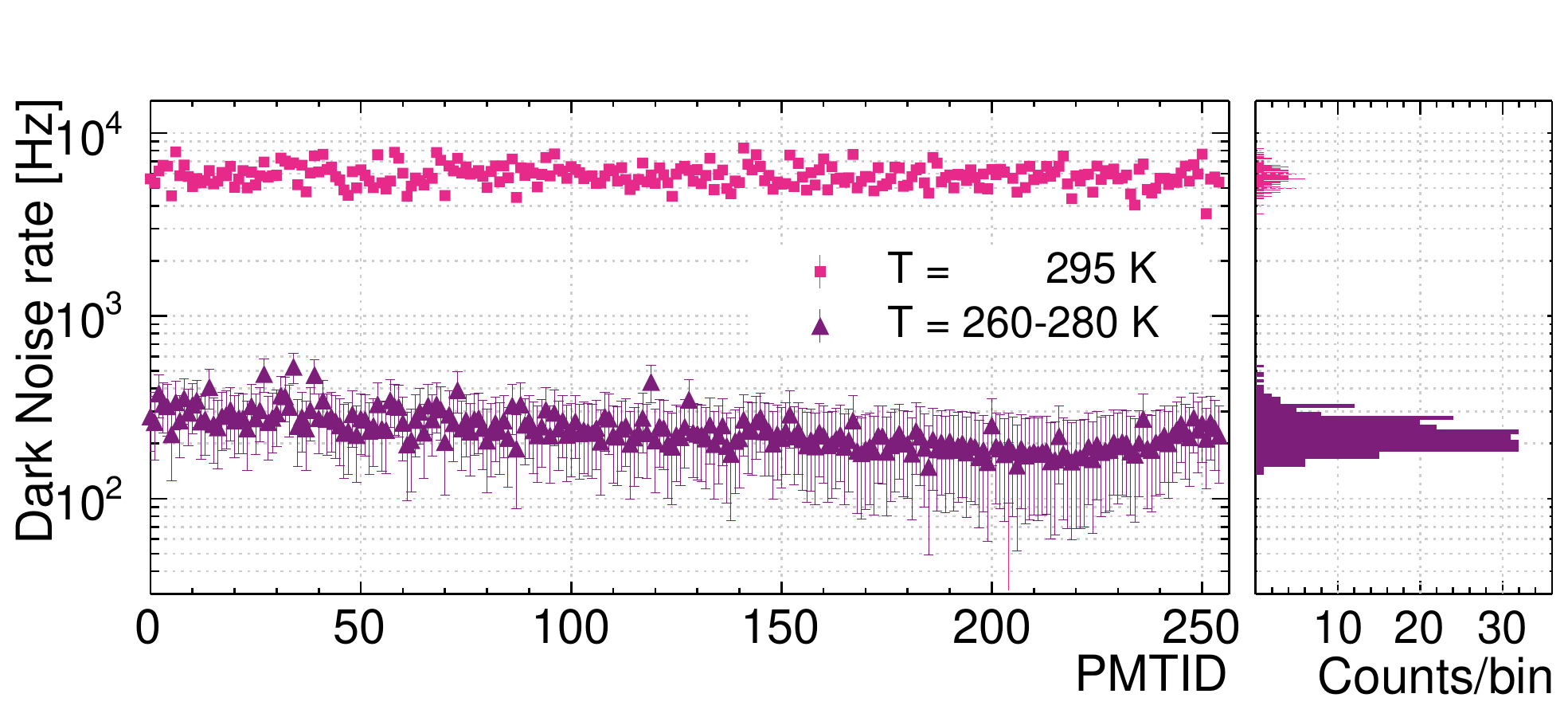}
	\includegraphics[width=3.5in]{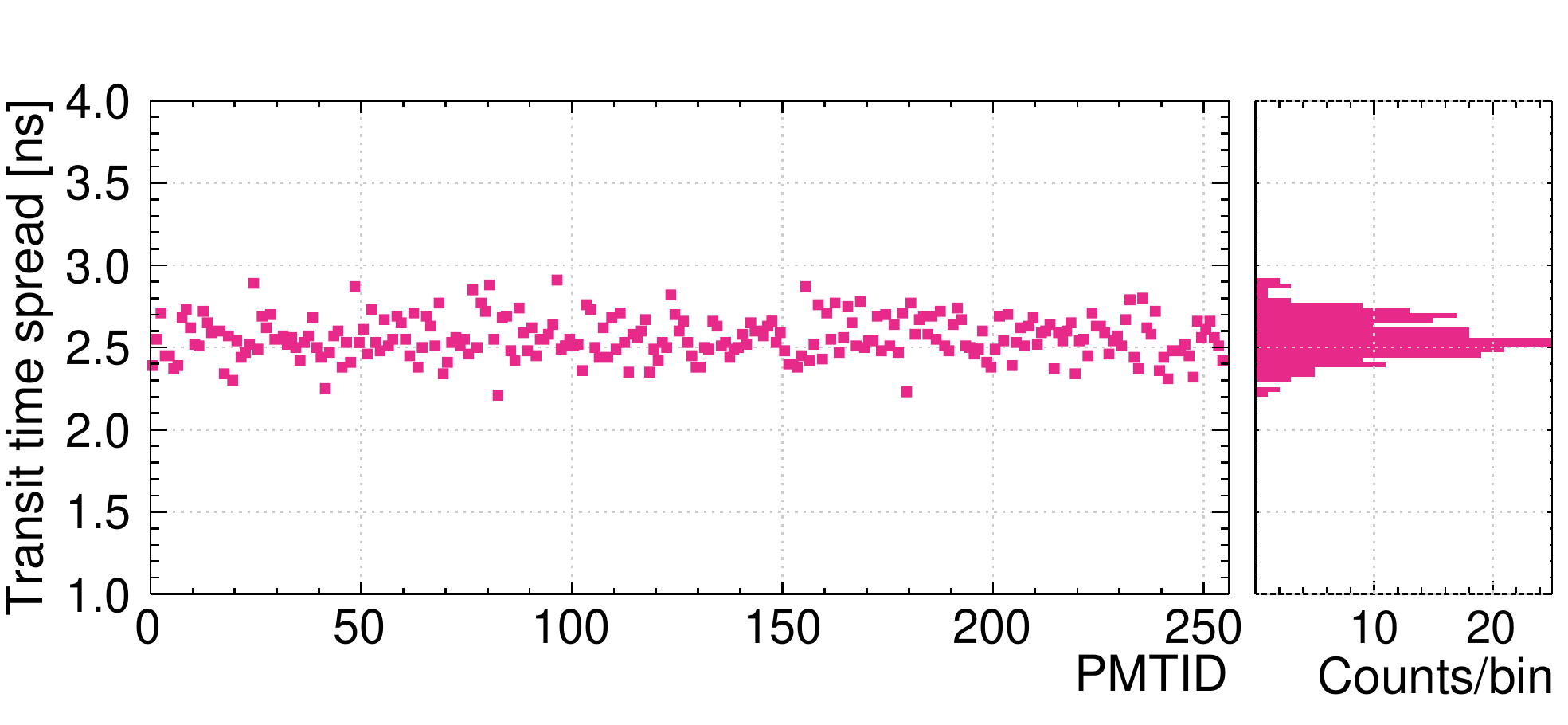}
\caption{Top: Dark noise vs.~PMTID for room-temperature (295~K, pink) and just after filling the detector with argon (260~K for large PMTID near the bottom of the detector and 280~K for small PMTID near the top of the detector. The room-temperature dark noise distribution has a mean of 5.8~kHz and an RMS of 0.78~kHz. The cold dark noise distribution has a mean of 0.24~kHz and an RMS of 0.06~kHz. Bottom: Full width at half maximum transit time spread vs.~PMTID, with a mean of 2.6~ns and an RMS of 0.12~ns. Error bars shown are statistical and typically smaller than the marker size.}
\label{fig:pmtchar2}
\end{center}
\end{figure}

Afterpulsing occurs in distinct time regions between 100~ns and 10~$\upmu$s and is caused when residual gas inside the PMT becomes ionized by moving electrons. The probability per PMT shown in Figure~\ref{fig:pmtchar:3} is the total probability of observing an afterpulse of any charge per SPE pulse within that time window. The relatively high afterpulsing probability has an mean of 7.1\% and an RMS of 1.8\%, which will be mitigated in analysis~\cite{pmtpaper}. 

\begin{figure}[h!]
\begin{center}		
	\includegraphics[width=3.5in]{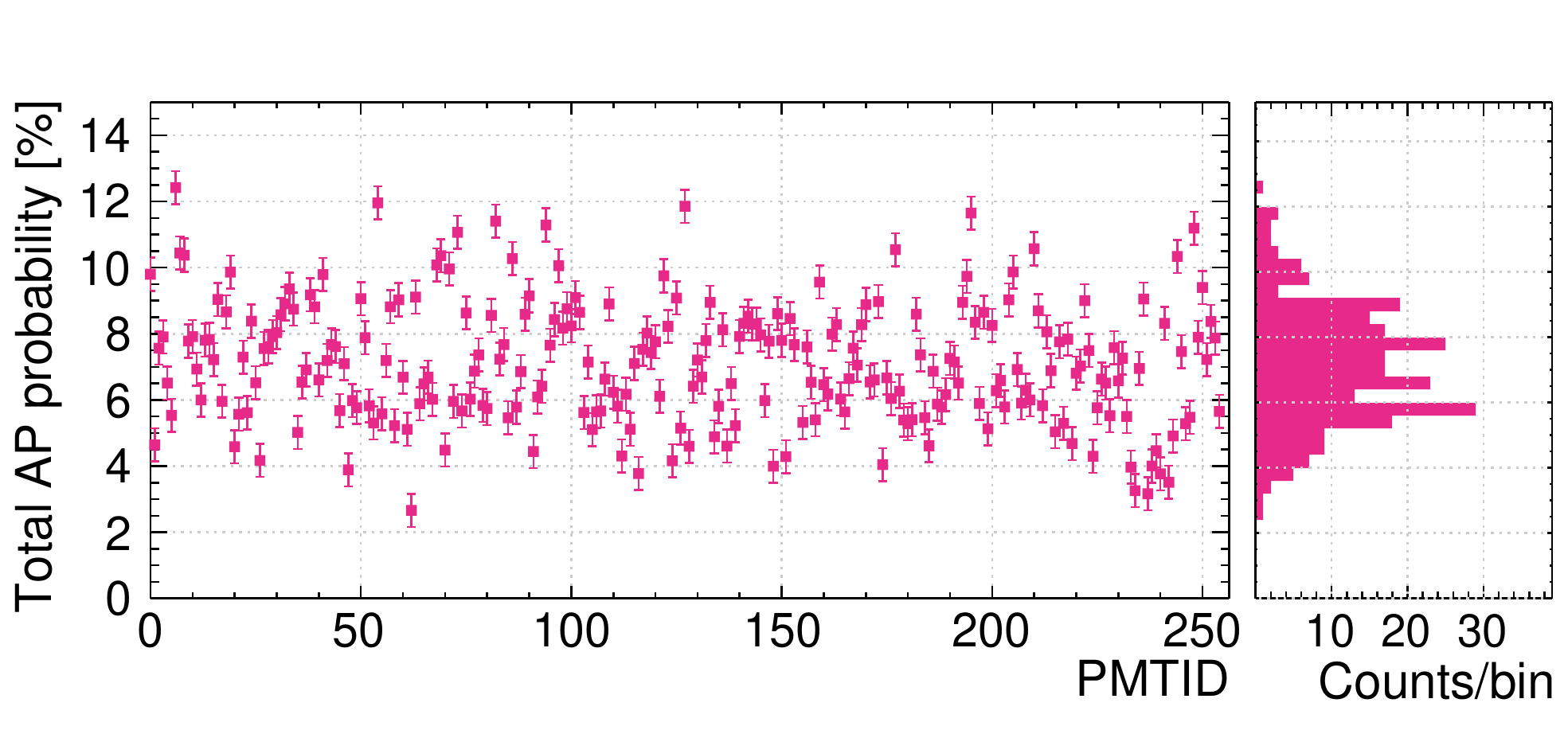}
\caption{Total afterpulsing probabilities vs.~PMTID. The distribution has a mean of 7.1\% and an RMS of 1.8\%.}
\label{fig:pmtchar:3}
\end{center}
\end{figure}

The double and late pulsing probabilities for the 255~PMTs are shown in Figure~\ref{fig:multipulsing}. In a double pulse, the full SPE charge is split over two separate pulses due to inelastic scattering of the photoelectron on the dynode. A late pulse is a similar effect, but in which the photoelectron backscatters from the first dynode without producing any secondary electrons. The distribution of double pulses has a mean of 2.7\% and RMS of 0.2\%, while the mean and RMS of the late pulse distribution is 2.3\% and 0.1\%, respectively. Characterization of these probabilities is used for time-resolved studies of pulse shapes at short times and time-based event position reconstruction. 

\begin{figure}[h!]
\begin{center}
	\includegraphics[width=3.5in]{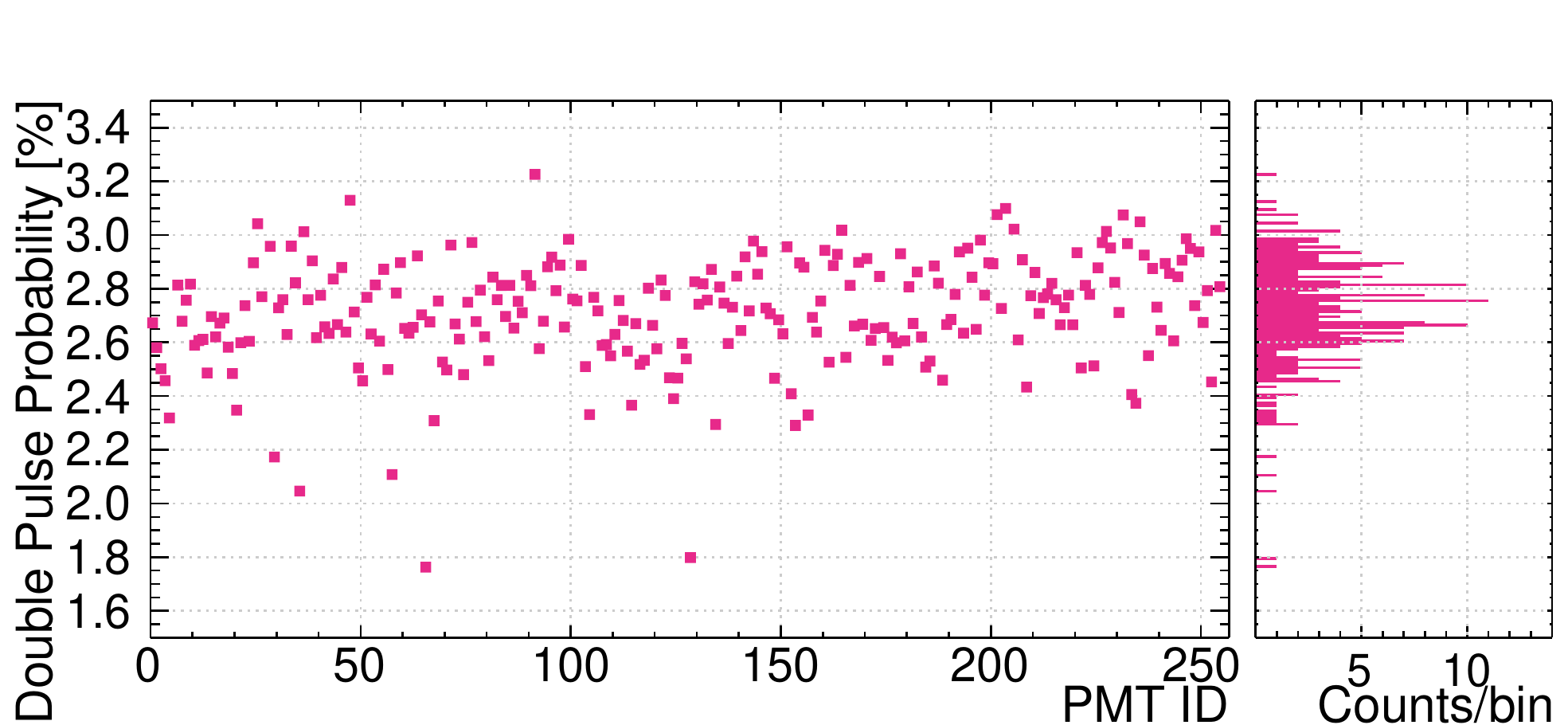}
	\includegraphics[width=3.5in]{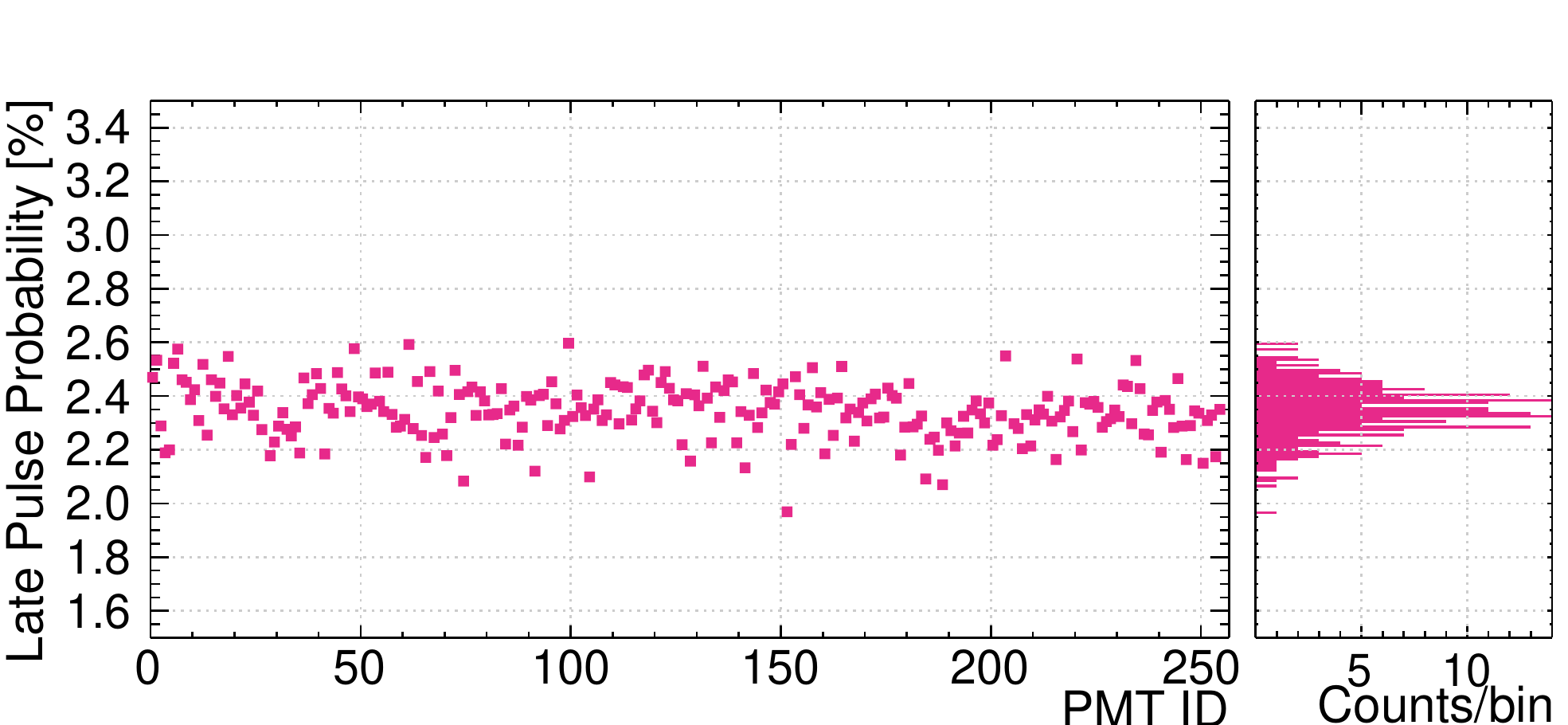}
	\caption[]{Top: Distribution of double pulse probabilities for the 255~inner PMTs vs.~PMTID with a mean of 2.7\% and RMS of 0.2\%. Bottom: Distribution of late pulse probabilities for the 255~inner PMTs vs.~PMTID with a mean of 2.3\% and an RMS of 0.1\%.}
\label{fig:multipulsing}
\end{center}
\end{figure}

\paragraph*{Magnetic Field Suppression}\label{subsec:magcomp}
A combination of field-compensating coils around the DEAP-3600 detector and an individual FINEMET shield around each PMT is used to mitigate external magnetic fields which can affect PMT efficiency and reduce gain. Although FINEMET saturates at a maximum flux density of 1.13~T, a small 60~$\upmu$m thickness has a maximum relative permeability of 70~kH/m~\cite{ref:finemet}, which makes it competitive with significantly thicker and heavier sheets of mu-metal.

Four compensation coils, each~11~feet in diameter, attached to threaded rods at $\pm~0.75$~m and $\pm~2.25$~m from the equator, are submerged in the water shield tank (Section~\ref{ssec:infrastructure}). This configuration was determined from a measurement of the ambient field in the detector surroundings using a hand-held geomagnetometer (Integrity Design IDR-321) and modeling with Radia~\cite{radia}. The coils are operated with remotely-controlled switching mode 600~W power supplies (Extech~382275). This arrangement leaves only an uncompensated 140~mG radial component, which is further reduced with the FINEMET shielding wrapping each PMT. Six 3-axis fluxgate magnetometers (Stefan Mayer Instruments FLC3-30~supplied with waterproof enclosures) were installed on the outside of the steel shell to monitor the field near the detector. These are read out through the DeltaV slow control system and logged for diagnostic purposes. 

\subsection{Neck Veto System}\label{ssec:neckveto}
The bottom 10~cm of the AV neck was wrapped with 100~Kuraray Y-11~(200M) wavelength shifting optical fibres, shown in Figure~\ref{fig:neckveto}, to provide additional information in identifying and discriminating backgrounds (e.g. Chereknov light) from this region of the detector where there may be incomplete light collection. Without full collection, a relatively high energy event could reconstruct with a low number of detected photoelectrons, potentially leaking into the energy ROI for the dark matter search. Each fibre is between 2.6~m and 3.3~m, and the fibre ends were gathered into bundles of 50 and optically coupled to four Hamamatsu HQE extended green R7600-300~PMTs. The PMTs were placed on top of the filler blocks near the neck at the same distance as the primary PMTs from the LAr volume. The neck veto PMTs are read out by the same DAQ hardware and software used for the inner detector PMTs. 

\begin{figure}[h!]
\centering
	\includegraphics[width=3.25in]{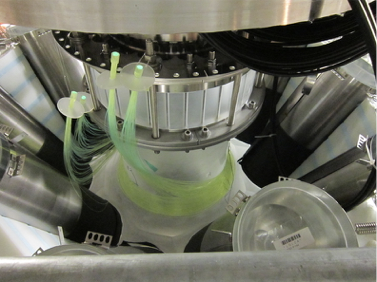}
\caption{The neck veto wavelength shifting fibers wrapping the bottom 10~cm of the AV neck before installation of the neck filler blocks. The fiber bundles were then optically coupled to R7600-300 PMTs (not shown).}
\label{fig:neckveto}
\end{figure}

\subsection{Calibration Systems}\label{ssec:calibration}
Calibration systems can be deployed in and around the detector to characterize the PMT response and event reconstruction. The optical calibration systems are comprised of a diffusing laserball source deployed inside the detector before it was cooled to characterize the optical response of detector materials, and an array of permanently-installed fiber optic cables, coupled to LED-drivers, to monitor PMT gains and pedestals as a function of time. All external radioactive calibration sources are deployed outside of the steel shell. A tagged \Nuc{Na}{22} gamma source is used to monitor the energy scale, resolution and position response, and a tagged americium-beryllium (AmBe) neutron source is used to monitor the response to neutron-induced nuclear recoils as expected from WIMP interactions. An untagged \Nuc{U}{232} source (15.6~kBq in March 2017) was deployed to study the detector response to gamma interactions in the acrylic producing Cherenkov light. Additionally, the intrinsic radioactivity from \Nuc{Ar}{39} beta decays provides a source of uniformly-distributed events with a known energy spectrum used to calibrate energy and position reconstruction biases from 50--5000 photoelectrons. 

\subsubsection{Optical Calibration}
\label{subsection:optical_calibration}
The initial optical calibration of the detector was performed with a nearly-isotropic optical photon source deployed once into the inner detector volume before the LAr fill. The laser-driven diffuse light source was based on the calibration source from the SNO experiment~\cite{ref:calibration_laserball}. A PerFluoroAlkoxy (PFA) flask, 11~cm in diameter, was loaded with 50-$\upmu$m quartz beads suspended in Silicone RTV-2~gel (Wacker Silgel 612 A/B) to produce a diffusing medium. Three laser diode heads with wavelengths of 375~nm, 405~nm, and 445~nm injected light from a Hamamatsu PLP-10 picosecond light pulser (70~ps typical FWHM) through a 1-mm-diameter optical fibre (Mitsubishi Rayon Co., LTD. SH-4001) into an acrylic stub with an end face centered in the PFA flask. These wavelengths were chosen to be below, near, and above the excitation wavelength of TPB; the highest wavelength has sensitivity only to the acrylic optical properties while the lowest wavelength additionally has sensitivity to the TPB coating optical response. The fast pulse additionally allowed for a measurement of the channel-by-channel timing offsets. The source was deployed in the AV using the same deployment system used for the TPB deposition to the center of the detector (Z=0) and $\pm$~55~cm, along with four $90^{\circ}$ azimuthal rotations at each z-position to disentangle non-uniformity in the source itself. The uniformity of the source was additionally measured using a CCD camera, ex-situ, to be within $\pm$~10\% with a 5\% measurement uncertainty. 

To calibrate the detector response as a function of time, optical calibration is performed with a permanently-installed LED light injection system. There are 22~light injection points in the detector, with 20~uniformly spread across the PMT array, and 2~located in the neck region. Additionally, there is one injection point in the water shield tank to calibrate the muon veto PMTs. At each point for the PMT array, a cylindrical section made from acrylic and coated in aluminum is attached to the LG perpendicular to a PMT face. Fast electronics based on~\cite{Kapustinsky1985} drive 435~nm LEDs to generate the light injected into the fibres, and an avalanche photodiode is used to monitor and correct for variations in LED intensity on a run-by-run basis.  Injected light travels through a 1-mm-diameter acrylic fibre to the cylindrical section, where it is reflected onto the PMT face. The LED-calibration PMT detects the light, and approximately 20\% is scattered down the LG into the inner detector volume, providing a diffuse light source to all PMTs more than approximately 50~degrees away from the active LED. The distribution of light from this calibration source can be seen in Figure~\ref{fig:optical_calibration_aarfsPC}, where the occupancy is defined as the fraction of the total light flashes that are registered in a given PMT.

\begin{figure}[h]
        \centering
        \includegraphics[width=0.75\textwidth]{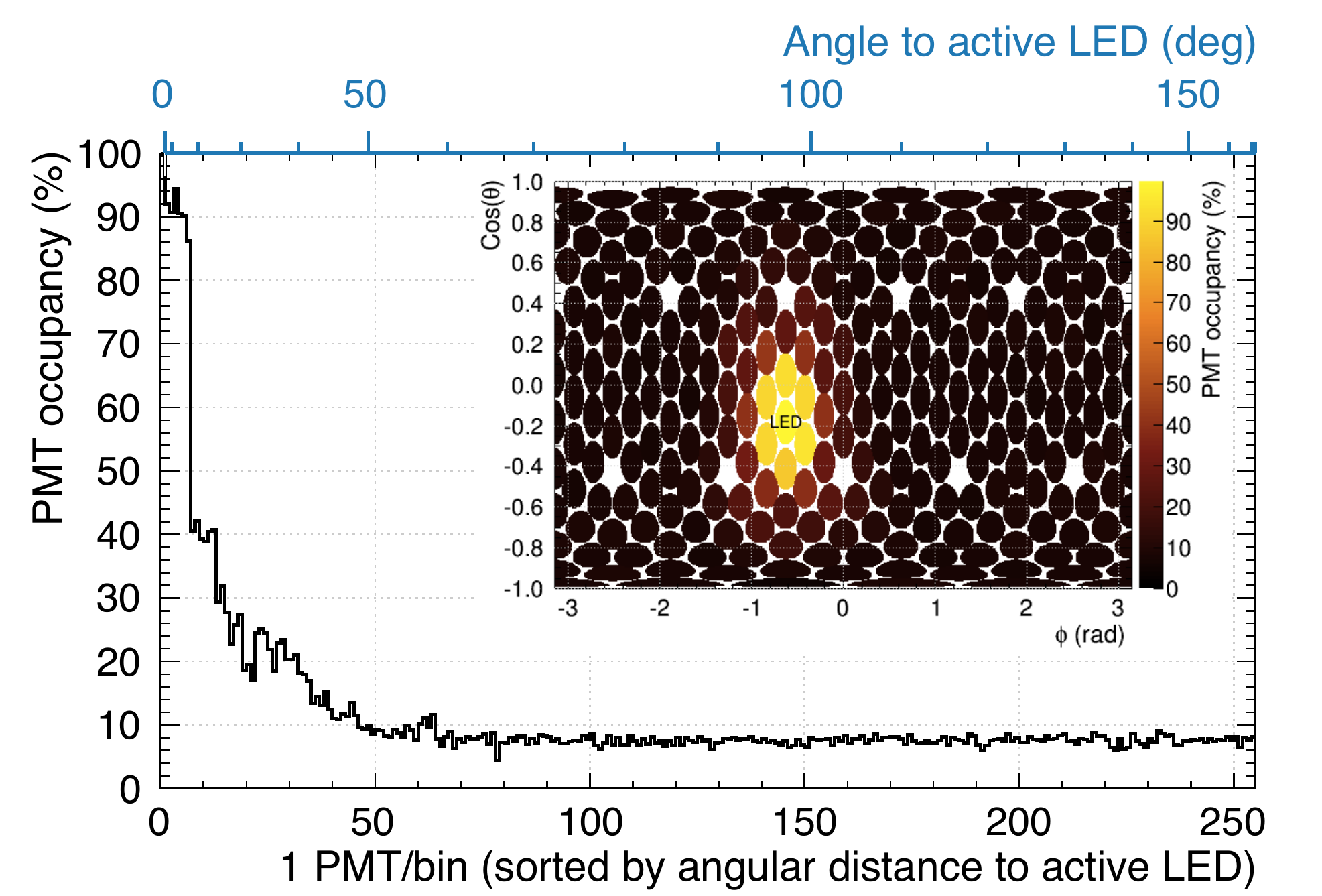}
\caption{Observed PMT occupancies from an LED calibration run vs.~PMT sorted by distance. The outer plot shows the average occupancy for each PMT, where the PMTs are sorted by ascending angular distance to the active LED. The inlay plot shows the PMT face positions projected onto a plane and colored by their occupancy. The active LED is marked as `LED' on the inlay plot. At an angular distance larger than 50 degrees from the active LED, the LED system provides a diffuse source of light.}
\label{fig:optical_calibration_aarfsPC}
\end{figure}

A measurement of the the relative PMT channel efficiency was performed using the diffuse laserball and LED calibration system. The relative efficiency is defined with respect to a single PMT channel. A strong correlation between the efficiency measured with each source is shown in Figure~\ref{fig:pmtEfficiency}. The RMS spread of the PMT channel efficiencies is measured to be 3.5\%.

\begin{figure}[h]
\centering
\includegraphics[width=4in]{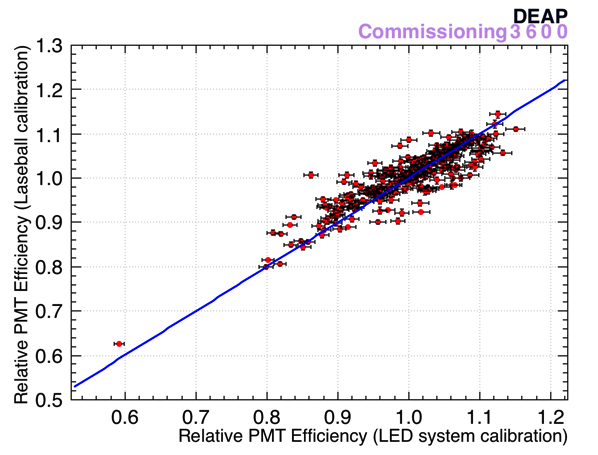}
\caption{Correlation between relative PMT channel efficiency measured with the laser optical calibration source vs.~the relative efficiency measured with the LED calibration system. The blue line represents the perfect correlation scenario. The two independent measurements agree within 7\%.}
\label{fig:pmtEfficiency}
\end{figure}

\subsubsection{Radioactive Calibration}\label{sssec:source_calibration}
There are 3~vertical stainless steel tubes (A, B (not visible), and E in Figure~\ref{fig:gamma_calibration_hardware}) used to deploy the tagged gamma and neutron calibration sources around the detector equator and one stainless steel tube (C in Figure~\ref{fig:gamma_calibration_hardware}) projecting outward from the upper hemisphere. A circular high density polyethylene tube (F in Figure~\ref{fig:gamma_calibration_hardware}) wrapping around the detector is also used with the gamma sources.

\begin{figure}[h]
        \centering
        \includegraphics[width=5in]{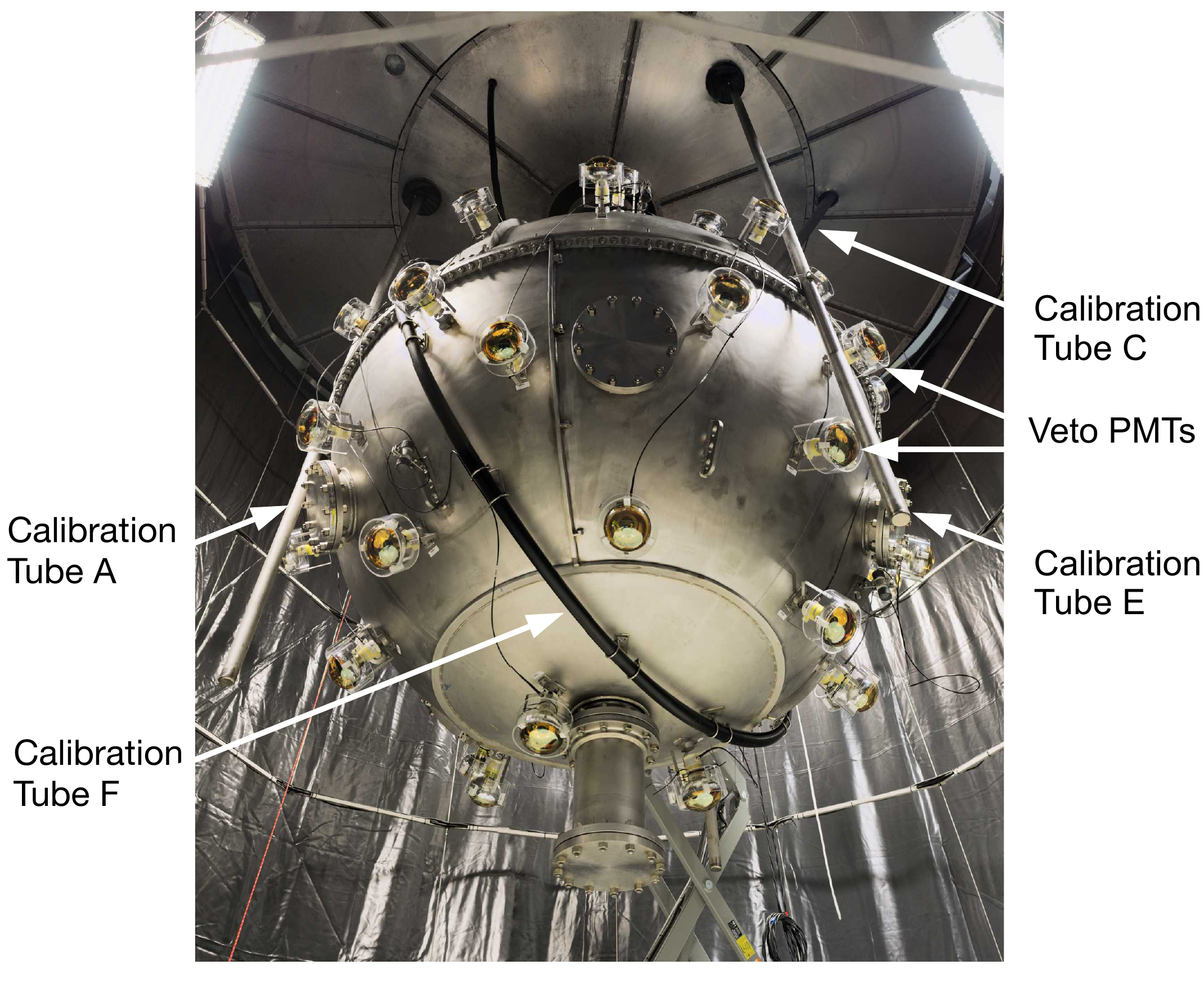}
        \caption{The DEAP-3600 detector in the water shield tank with the vertical calibration tubes A, B (not visible; in background), C, E, and circular high density polyethylene tube, F, indicated. The 48~PMTs attached to the steel shell are used for the muon veto.}
\label{fig:gamma_calibration_hardware}
\end{figure}

A pulley and carriage system is used to deploy the gamma and neutron source canisters, and is driven by a Mclennan 34HSX-208E stepper motor through a Mclennan SimStep controller. The deployable source position uncertainty was measured ex-situ to be approximately 1~cm. 

The 1~MBq \Nuc{Na}{22} source (created April 2012) is contained between two Cerium-doped Lutetium Yttrium Orthosilicate (LYSO, Hilger~\cite{ref:calibration_crystals}) crystals to tag the back-to-back 511~keV annihilation photons. The 20-mm-diameter, 20-mm-long LYSO crystals are read out with two compact Hamamatsu R9880U PMTs~\cite{ref:calibration_hamamatsu}. The tagging system can be compact, as 50\% of the 511~keV photons are attenuated in 8.5~mm in LYSO. 

The 74~MBq~AmBe neutron source is surrounded by two Ametek-packaged NaI crystal and PMT assemblies. Each 40-mm-diameter, 51-mm-thick NaI crystal has a 10~mm by 10~mm well to contain the radioactive source, and is coupled to a 38~mm ETL 9102~PMT with an internal Cockcroft-Walton high voltage generator allowing the PMT voltage to be driven from a 5~V source. 

Neutrons are created when an alpha produced by the decay of \Nuc{Am}{241} captures on a \Nuc{Be}{9} nucleus releasing a neutron and a 4.4~MeV gamma. The tagging PMT thresholds are set such that the gamma is used as a tag for the emitted neutron. The AmBe source is wrapped in 2~mm of lead foil to remove 99.9\% of the 60~keV gammas produced in addition to the alphas from the decay \Nuc{Am}{241}. The neutron rate from the source, as measured by the manufacturer (Eckert and Ziegler Isotope Products), is 4.8 kHz. 
\section{Electronics}\label{sec:electronics}
The overall architecture of the data acquisition (DAQ) system is shown in Figure~\ref{fig:daqConcept}. PMT signals are analyzed by the Digitizer and Trigger Module (DTM), which decides whether to trigger event readout. Trigger signals are sent to commercial digitizers (CAEN V1720s~\cite{v1720_man} and V1740s~\cite{v1740_man}), which digitize the PMT information. The digitized information is then read out, filtered, and written to disk.

\begin{figure}[h!]
\begin{center}
\includegraphics[width=\textwidth]{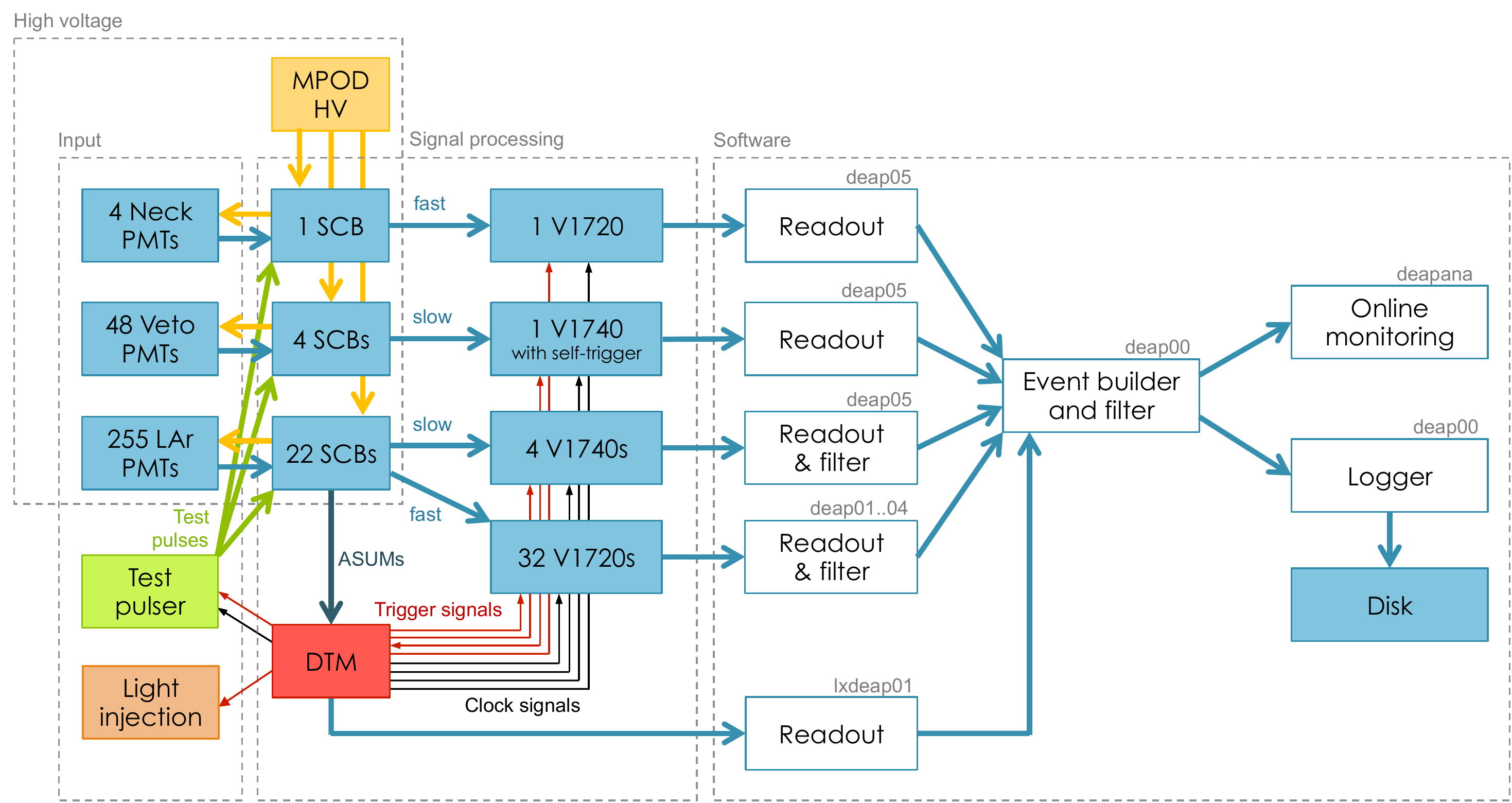}
\caption{Overall DEAP electronics architecture. Shaded boxes are hardware components; white boxes are software programs. SCBs are signal conditioning boards to shape PMT signals. V1720s and V1740s are commercial CAEN digitizers. DTM is the trigger module.}
\label{fig:daqConcept}
\end{center}
\end{figure}

\subsection{Hardware}\label{ssec:daqHardware} 

The DAQ system is housed in 3~computer racks. Each rack has an uninterruptible power supply with enough power to safely shut down the system in case of a power outage underground.

\subsubsection{PMTs and Signal Conditioning}

In addition to the 255~signal PMTs described in Section~\ref{ssec:PMT}, signals from the 48 Hamamatsu R1408 8-inch PMTs (45 active) for the outer detector muon veto, shown in Figure~\ref{fig:gamma_calibration_hardware}, and 4 Hamamatsu R7600-300 PMTs forming a neck veto (see Section~\ref{ssec:neckveto}) are recorded by the DAQ. 

The PMTs are powered by a WIENER MPOD crate~\cite{mpod} with ISEG high voltage modules~\cite{iseg} through 27~custom Signal Conditioning Boards (SCBs). Each SCB handles up to 12~PMTs, with 22~boards dedicated to inner-detector PMTs, 4~to the muon veto PMTs, and 1~to the neck veto PMTs. The SCBs decouple the high voltage, provide high voltage protection, and shape the PMT signals. The PMT bases are back-terminated with a 4.7~nF capacitor in series with a $75~\Omega$ resistor, and connected to the SCBs by a $75~\Omega$ impedance cable, each approximately 20~metres long. 

Each SCB has 12~identical channels to shape and amplify the PMT signals. There are three outputs from each channel: a high-gain channel, a low-gain channel, and a summing channel. The high-gain channel is designed to achieve high signal-to-noise for single photoelectrons and shape the pulse to better match the 250~MS/s V1720 digitizer. The low-gain channel is designed to handle pulses that saturate the high-gain channel and is attenuated by a factor 10~in amplitude. The low-gain pulses are also shaped to be significantly wider, to better match the 62.5~MS/s V1740 digitizer. The 12~summing channels are added to create an analog sum (ASUM) for each SCB. The 22~ASUMs from inner detector SCBs are passed to the DTM with a 24-channel differential connector.

In addition to the 12~safe high voltage (SHV) inputs from the PMTs, each SCB contains a ``test pulse" input. The test pulse is created by the DTM, and is sent through a discriminator and a fan-out board, to be distributed to all SCBs simultaneously. Within each SCB, the test pulse is distributed to all 12~channels, with a 0.2~ns channel-to-channel delay. This system allows for easy extraction of timing offsets between different digitizer channels.

Tagging PMTs for calibration sources are also read out by the DAQ when deployed. These PMTs are powered by external power supplies, rather than the SCBs.

\subsubsection{Digitizers}

The high-gain outputs from the SCBs are connected to 250~MS/s CAEN V1720 waveform digitizers (8~channels, 12~bits) using MCX cables. V1720s can store data either in Zero Length Encoding (ZLE) mode or as full waveforms. An example of an SPE pulse in ZLE mode is shown in Figure~\ref{fig:v1720_spe}. The ZLE algorithm records data only if a given number of samples drop below a threshold ADC value. The noise level is approximately 1.2~ADC on the V1720 channels, and a typical SPE pulse is approximately 50~ADC high. The ZLE threshold is set to be 5~ADC below the baseline of 3900~ADC, providing a balance between recording real pulses and limiting recorded noise fluctuations. Additionally, 20~extra samples (80~ns) before and after the pulse are recorded. The baseline is set to 3900~ADC, rather than closer to the maximum 4096~ADC, to allow overshoot to be recorded. 

\begin{figure}
\centering
\includegraphics[width=3.0in]{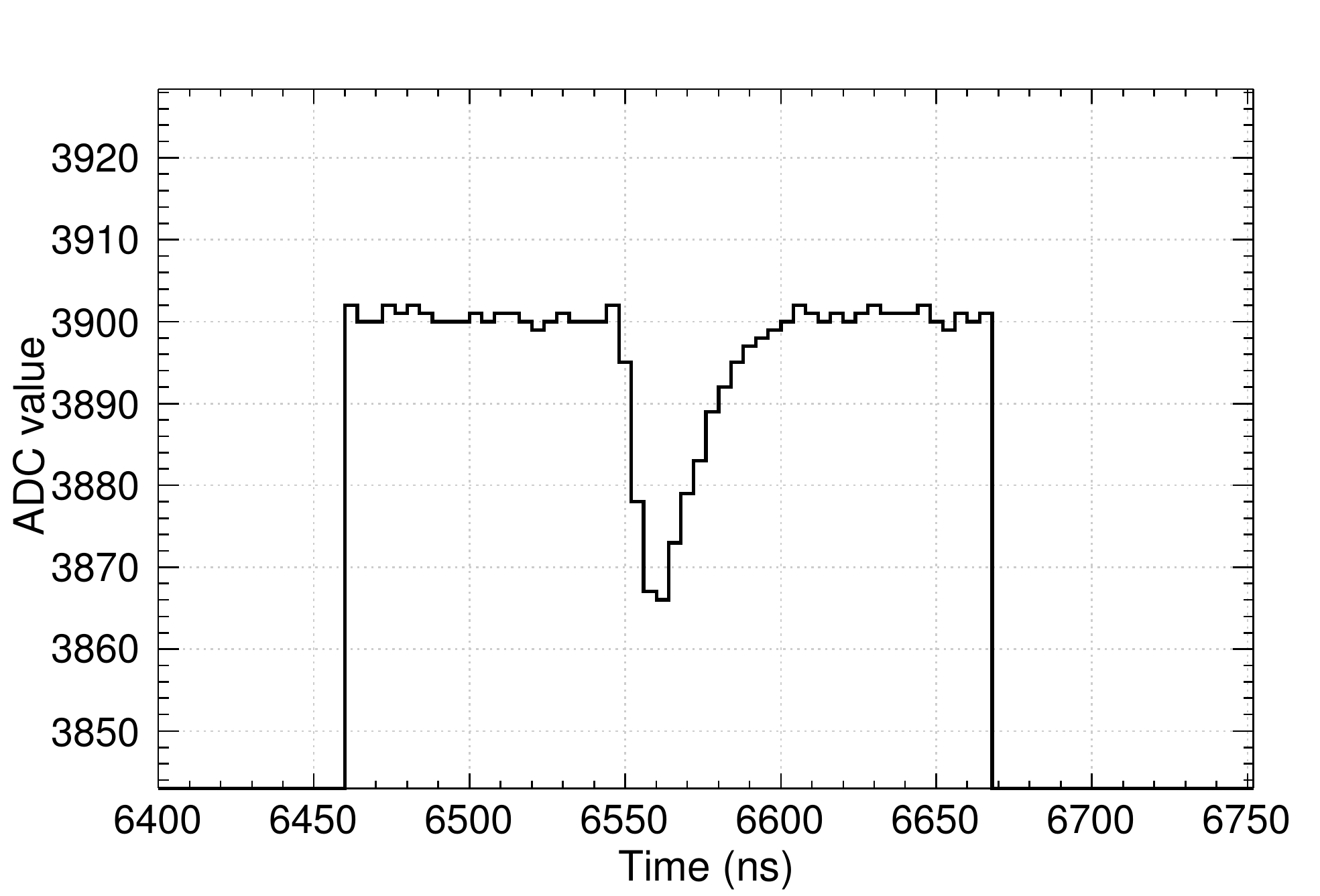}
\caption{Example of a measured SPE pulse on a V1720 channel, in ZLE mode. The baseline is set to 3900~ADC, the ZLE threshold for saving data is 3895~ADC, and 20 extra pre-samples and post-samples are saved. Time is in 4~ns binning.}
\label{fig:v1720_spe}
\end{figure}

The low-gain outputs from the SCBs are connected to 62.5~MS/s CAEN V1740 waveform digitizers (64~channels, 12~bits) using MCX cables. The V1740s do not allow for recording data in ZLE mode, and can only record full waveforms. Software is used to filter these waveforms so they do not dominate the data rate (see Section~\ref{ssec:daqSoftware}). 

The muon veto PMTs are connected to a V1740~running in ``self-trigger" mode. Instead of the DTM analyzing the veto PMT signals, the digitizer itself decides whether there is sufficient activity in the water tank to trigger readout. The 48~PMTs are separated into 6~groups of 8~PMTs, and if any channel in a group exceeds a height threshold of 15~ADC (approximately 0.75~PE), that group is deemed to be ``active". If three groups are simultaneously active, then the self-trigger condition is met. A signal is then sent to the DTM.

The neck veto PMTs are connected to a V1720~running in ZLE mode. During calibrations with radioactive sources that have tagging PMTs, the signals from the tagging PMTs are also handled by this V1720.

The digitizers are read out through optical links using proprietary CAEN A3818~cards~\cite{a3818_man}. Each card handles 4~optical links, and two V1720s are daisy-chained on the same link. Each card therefore reads out 8~V1720s, or 4~V1740s.

\subsubsection{Trigger}

The DTM is responsible for making the trigger decision, providing the master clock to synchronize digitizers, triggering digitizers and external calibration systems, and throttling data collection if the DAQ is overwhelmed. The DTM hardware is based on a TRIUMF-designed 6U VME motherboard with an ALTERA Stratix IV GX field-programmable gate array (FPGA). The motherboard has three daughterboards connected through FMC standard connectors~\cite{fmc_standard}. The daughterboards are a 24-channel ADC card for digitizing the ASUM channels from the SCBs~\cite{trig_adc}, a 12-channel NIM I/O card with 8~outputs and 4~inputs, and a master clock distribution board. The master clock is distributed at 62.5~MHz to the digitizers, while the ADCs and main FPGA run at 45~MHz. The NIM outputs are connected to the digitizers, LED light injection system (Section~\ref{subsection:optical_calibration}), and a test pulse system.

The trigger system is based on a set of ``trigger sources" and logical ``trigger outputs". Trigger sources, from internal analysis of PMT signals, periodic triggers, or external triggers, are responsible for deciding whether a trigger signal should be issued. Trigger outputs decide which hardware should be fired. A logical ``trigger output" may fire one, several, or no NIM outputs, and can be configured to ignore a certain percentage of trigger signals (pre-scaling). Each trigger source can be connected to many trigger outputs, all configured differently. 

The main trigger algorithm used in DEAP-3600 is the ``physics trigger", which adds the 22~ASUMs together for a sum of all 255~inner-detector PMTs. Rolling integrals are computed in two windows, nominally 177~ns and 3100~ns, aligned to the same start time. The total charge in the prompt window ($E_\mathrm{prompt}$) and the ratio of energy in the prompt and late windows ($F_\mathrm{prompt}$) are calculated. The $(E_\mathrm{prompt}, F_\mathrm{prompt})$ phase space is split into 6~regions, as shown in Figure~\ref{fig:triggerRegions}. Events in region X are discarded, while the other 5~regions all count as separate trigger sources. The $E_\mathrm{prompt}$ threshold of 1000~ADC gives a 50\% detection efficiency for events with 30 prompt PE. The lower bound of region E is set to be above the end of the \Nuc{Ar}{39} beta decay spectrum.

\begin{figure}
\centering
\includegraphics[width=3.0in]{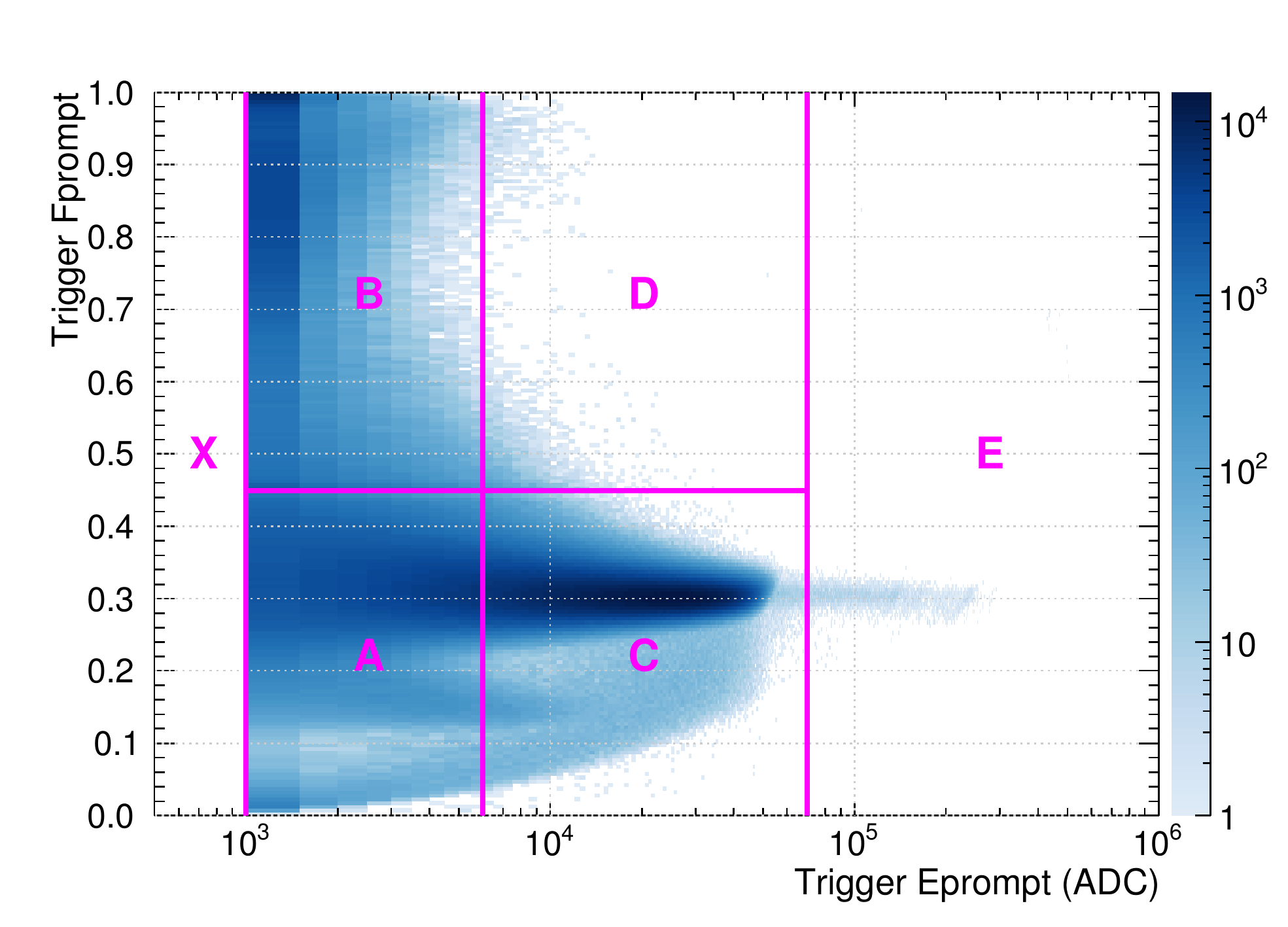}
\caption{Trigger prompt energy and trigger $F_\mathrm{prompt}$ for example data from the physics trigger, with no cuts applied. Darker colours indicate more events. The 6~labelled trigger regions are shown in magenta.}
\label{fig:triggerRegions}
\end{figure}

The standard trigger setup for DEAP-3600 data-taking uses the physics trigger, a periodic trigger, and the muon veto self-trigger. The physics trigger is set to not read out the digitizers for 99\% of events in region C, which is dominated by \Nuc{Ar}{39} beta decays. Digitizers are read out for all events in regions A, B, D and E. Summary information (time, $E_\mathrm{prompt}$ and $F_\mathrm{prompt}$) is stored for all events, regardless of whether the digitizers were read out or not. The periodic trigger runs at 40~Hz, with test pulses injected at 1~Hz. The remaining 39~Hz are used to monitor the PMTs, as described in Section~\ref{ssec:daqOperation}. The veto PMTs are only read out when the muon veto self-trigger fires. The overall trigger rate is 3200~Hz, with digitizers read out at 500~Hz.

An additional trigger mode is for daily LED light injection calibration, which only uses periodic triggers. The monitoring trigger is accompanied by a 1~kHz periodic trigger that fires the light injection system and digitizers.

\subsection{Software and Data Rate Reduction}\label{ssec:daqSoftware} 

The readout software is responsible for interfacing with the digitizers and the DTM, filtering out unnecessary information, collecting information into a single event, and writing that event to disk. In total, 7~computers are involved in the readout, all running Scientific Linux 6.6. Four computers handle the inner detector V1720 data, with additional PCs for the DTM data, the inner detector V1740 data, the muon veto and neck veto data, and a master.

The V1720 digitizer saturates at approximately 100~PE, and V1740 information is only needed for pulses larger than this. Unnecessary V1740 data is filtered out in two stages. The first stage only relies on V1740 information, and filters out any waveforms that do not go below 3750~ADC, from a baseline of 3900~ADC. This reduces the V1740 data to be passed to the master computer, where the second filtering stage removes V1740 waveforms in which the the corresponding V1720~channel does not go below 500~ADC. Over 99.9\% of V1740 information is removed through this process.

Filtering is also applied to the V1720 waveforms, to further reduce the amount of data written to disk, and only summary information about the pulse is saved. This summary information is sufficient to give sub-ns timing resolution of the peak position, as well as the pulse charge, height, and baseline (with RMS). SPE identification uses probability distribution functions of the ratio of pulse height to pulse charge, the ratio of pulse height to maximum derivative, the width, and total charge of the pulse.

After all filtering and compression measures, the data rate is reduced from 7~GB/s to 6~MB/s.

\subsection{Operation}\label{ssec:daqOperation} 
The DAQ system is based on the MIDAS package~\cite{midas_ref}, developed by TRIUMF and the Paul Scherrer Institute. A web interface has been written to allow easy remote usage of the DAQ which additionally interfaces with DEAP's CouchDB database. The database stores DAQ parameters to be used for given run types, which are selected by the operator at the start of each run and automatically forwarded to the appropriate DAQ components. 

The DAQ has been designed to run semi-autonomously, with fail-safes in case of software, hardware or network malfunction. In these scenarios the current run is stopped and an SMS and email are sent to the DAQ operator. If the operator does not fix the problem and start a new run within 15~minutes, the PMTs are ramped down. The entire DAQ system performs a controlled shutdown in case of power loss or excessive temperature in the DAQ racks due to loss of cooling water. The data-taking uptime is greater than 95\%, excluding unexpected power outages.

\subsection{Database and Data Flow}\label{ssec:dataFlow} 

DEAP uses an Apache CouchDB database~\cite{CouchDB} to consolidate all external parameters necessary to configure the DAQ, to calibrate and analyze detector data, and to evaluate the data quality. Parameter validity ranges are implemented by run number, and parameter values are version-controlled. Additionally, readings from the DeltaV slow controls system, which monitors physical detector statuses that are relevant to the analysis, are transferred continuously to a PostgresSQL database. The DEAP analysis framework transparently queries either the CouchDB or the PostgresSQL database in support of analysis tasks.

ROOT files containing raw detector data are transferred from the DAQ storage computer to the main analysis cluster. A first-pass analysis translates DAQ into physics units, applying all necessary calibrations and corrections such as PMT gains and channel timing offsets. Simultaneously, diagnostic plots for data quality control are generated, which are available online at the end of the run. The second analysis pass starts from the calibrated data and calculates high level information about each event, including event position and number of detected photoelectrons. The result of each reduction step is written to a separate set of ROOT files.

The analysis software is maintained in a Git repository on a self-hosted GitLab server, which provides a web interface and performs automated builds and tests after every commit (GitLab-CI).
\section{Detector Infrastructure}
\label{ssec:infrastructure}
The DEAP-3600 detector is located in the Cube Hall at SNOLAB next to the MiniCLEAN experiment~\cite{rielage2015update}. A schematic drawing of the detector within the larger infrastructure is shown in Figure~\ref{fig:InfrastructureFromModel}. The Cube Hall has approximately $15 \times 15 \times 15$ m$^{3}$ of useable space and is furnished with an overhead 9-tonne monorail crane and a 10-tonne gantry crane (Konecranes) on the deck. 

\begin{figure}[h!] 
\hspace{-5mm}
     \includegraphics[width=5.in]{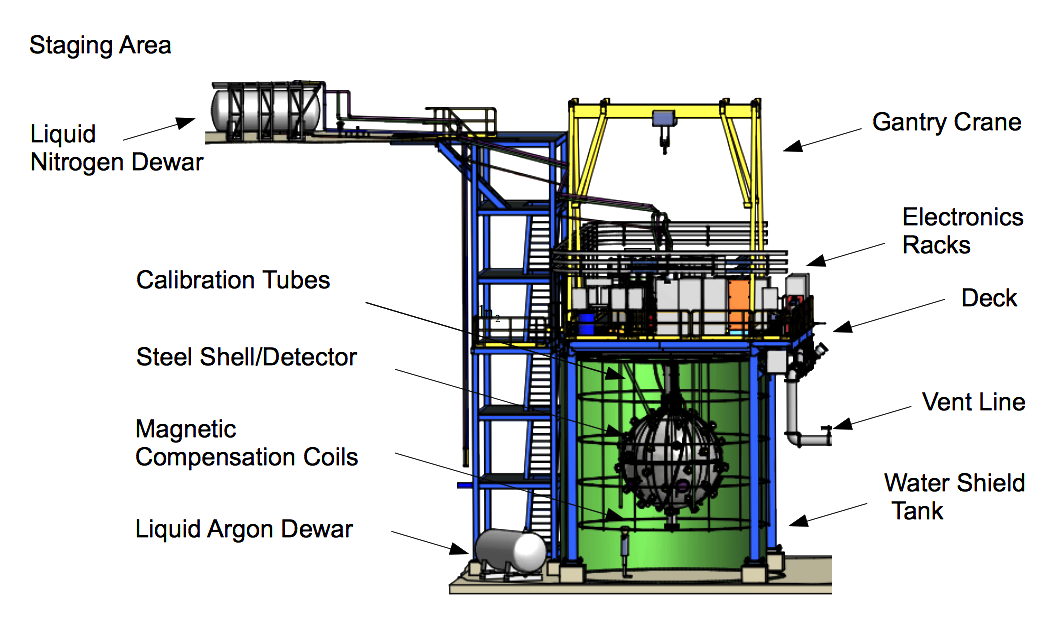}
     \caption{Model rendering of the DEAP-3600 detector and surrounding infrastructure inside the Cube Hall at SNOLAB.}
     \label{fig:InfrastructureFromModel}
\end{figure}

The deck is supported by 6~columns and stands over the two experiments to support the detector, process systems hardware, and electronics. The detector is housed in a spherical steel shell hanging from a 45-cm-diameter outer neck rigidly coupled to the central support assembly (CSA). The CSA is connected through seismic bushings to the deck to allow movement during a seismic event. The system is designed for survival of the SNOLAB seismic design event, a Nuttli 4.3~seismic event at 160~m from the site with a peak particle velocity of 800~mm/s~\cite{muonSNOLAB}. 

The steel shell, 3.4~m in diameter and fabricated from 304~stainless steel (All-Weld, Toronto, Canada) with an electropolished inner surface acts as a water-tight and light-tight vessel, and directs cryogen boil-off to a vent header in the event of a structural failure in the AV. It was designed as an ASME Section VIII pressure vessel to withstand the maximum pressure of 30~psig which could be developed with a fracture in the AV. The entire steel shell was helium leak tested after construction. 

Hanging coaxially inside the steel shell neck is a 30-cm-diameter stainless steel inner neck 3~m in length, which is coupled to the CSA through a custom load cell (Sensing Systems) and supports the approximately 13,000~kg inner detector (LAr, AV, LGs, filler blocks, PMTs and detector cabling). The inner neck was fabricated from a single length of seamless tubing and electropolished. Detector cabling and gas purge lines feed along the outside of the inner neck, and pass through vacuum feedthrough flanges on the CSA. The vapor space between the inner and outer necks provides the relief path for boil-off argon gas in the AV failure scenario. The AV bolts to the bottom of the inner neck through an acrylic-to-steel coupling flange. When cold, the acrylic flange contracts significantly more than the metal. To compensate for this, Belleville Spring washers were installed to ensure the bolts maintain sufficient clamping force through cooling.

The glovebox, shown in Figure~\ref{fig:design}, is a permanent interface located on the deck at the top of the neck, used to deploy components to the interior of the detector in a clean, radon-reduced environment. It is a 76-cm-diameter, 76-cm-tall cylindrical stainless steel vessel maintained either under vacuum or with a radon-scrubbed boil-off nitrogen over-pressure purge. The glovebox is equipped with butyl rubber dry-box gloves for access and inside manipulation. A 45-cm gatevalve on the top permits loading of large items. Additional 5.5-m and 4.8-m-long cylindrical deployment canisters would be bolted to the gatevalve, and with the use of an internal hoist were used to extend the clean glovebox volume for extraction of the resurfacer, and to execute the TPB deposition and diffuse-laser ball calibration, and to install the neck cooling coil. 

The steel shell is surrounded by a cylindrical water shield tank with a diameter and height of 7.8~m made from curved galvanized steel panels. A vinyl U-shaped liner is located inside the cylinder with an additional vinyl backing material between the inner liner and tank. A SQ26-10 high sensitivity hydrophone (Sensor Technology Ltd.) is installed in the water tank and read out with the DAQ to provide additional information on seismic events.

The water shield tank has a 75-L/min purification system consisting of a magnetically coupled pump (IWAKI America MX-251), mixed bed ion exchange columns (Purolite NRW-37), 0.45-$\upmu$m filtration (Shelco 3FOS2), degasser (Membrana Liqui-Cel 4x28) and a 254-nm UV sterilizer (Viqua UV Max F4 Plus). An optional flow path for degassing the water with a nitrogen purge is also provided. The purification system is run continuously. Two water chillers (Tek-Temp Instruments, Inc.) maintain the shield tank water at approximately 12$^{\circ}$C. Additional chilled water supplies are provided by SNOLAB for cooling of the DAQ, LN$_2$ cryocoolers, and as-needed auxiliary systems. 
\section{Safety}\label{sec:ODH}
Large volumes of cryogen are used underground for both the DEAP-3600 and MiniCLEAN experiments. For the safety of underground personnel, over-pressure protection and oxygen-deficiency analyses were performed. 

\subsection{Over-pressure Protection}
A failure of the DEAP AV would allow the LAr at 87~K to spill onto the warmer detector components, quickly boiling and releasing a large volume of inert gas which would lead to an over-pressurization of the steel shell. To mitigate this hazard, a burst disk was installed near the top of the steel shell neck feeding to a 12-inch-diameter vent pipe which services both DEAP-3600 and MiniCLEAN. It is 335~m long and terminates at a large-capacity mine-air raise. The 120,000 cfm upward air flow serves to dilute the vented gas and remove it from the mine. A detailed analytic model was developed to predict the worst-case vapor generation rate as a function of time. Sample room-temperature detector components were dipped into LAr and the measured vapor generation rates were then extrapolated to the full detector. This model predicts a peak vapor generation rate of 45 kg/s which decreases to less than 5 kg/s after 1~minute.  

In addition to the vent pipe, a secondary pressure safety valve provides pressure relief into the Cube Hall. An 8-inch-diameter pipe connects the CSA to a rupture disk set at 15~psig followed by a check valve set at 5~psig. The main vent pipe can provide a steady-state flow of 13~kg/s of argon gas with a pressure drop of 15~psig. An analytic model of the transient gas flow using the worst-case vapor generation rate predicts that the secondary pressure safety valve would open for approximately 80~s and release 220~kg of argon gas into the Cube Hall. 

\subsection{Oxygen Deficiency}
Analysis of the oxygen deficiency hazard (ODH) followed the methodology developed at Fermi National Accelerator Laboratory~\cite{fermilabodh} (Fermilab), which maintains tables of failure rates for many components used in cryogenic systems. 

Several techniques have been implemented to mitigate the oxygen deficiency hazard. The large cryogen storage vessels are certified ASME section VIII pressure vessels, resulting in a low probability of failure. Additionally, the outlet of the nitrogen storage vessel has a flow-restriction of 100~g/s. Seismic isolation pads are used for the argon dewar, nitrogen dewar, and steel shell, and mechanical barriers are in place to prevent contact with dewars by heavy equipment.

Due to the large volume of the Cube Hall, mixing of the air is sufficient to dilute vented gas in all scenarios. In addition to the main SNOLAB ventilation system, a secondary air mixing system was installed. This consisted of three 5000~cfm and two~of 300~cfm fans on the Cube Hall floor, two 300~cfm fans on the deck, and two 1000~cfm fans at the top of the Cube Hall near the nitrogen dewar. These run at all times and the supplied currents are monitored and backed up by an uninterruptible power supply. A series of oxygen monitors (PureAire TX-1100-DRA) are installed around the Cube Hall and in adjacent halls. 

Spill tests, in which 100~kg to 150~kg of argon were flash boiled on the Cube Hall floor in less than 30~seconds, generated oxygen levels below 135~mmHg, not safe for personnel, at the Cube Hall floor monitors. The oxygen levels were extrapolated to the model failure scenario. From this, a set of four 2500~cfm on-demand vertical mixing fans (Pearson 12~inch Velocity), triggered when the oxygen sensors read low, was installed to move air from the Cube Hall floor to the ceiling to promote mixing and dilution. 

With all mitigation steps in place, the probability of a serious accident using the Fermilab methodology is below $10^{-7}\,\mbox{hr}^{-1}$ which places the Cube Hall in a category that does not require staff to routinely have access to a re-breather, self-contained breathing apparatus, or other emergency air systems. 

\section{Summary}
The DEAP-3600 detector searches for dark matter particle interactions using single-phase liquid argon technology. The projected WIMP-nucleon cross-section sensitivity for a 3-tonne-year fiducial exposure is $10^{-46}~\rm{cm^{2}}$ at 100~GeV/$c^2$ WIMP mass. 

The use of a simple, single-phase liquid argon target contained in an acrylic cryostat is novel. Significant design, research, and development have been undertaken to minimize detector backgrounds including quality control during the ultra-pure acrylic vessel manufacturing and resurfacing, the selection of low radioactivity materials, and limiting the exposure of detector components to radon during assembly and construction.

The PMT system has been operational since the end of 2014. The cryogenic handling and purification system was commissioned in winter 2014. The water shield tank components, application of the wavelength shifter, calibration hardware and muon veto PMT system were completed in summer 2015. Installation of the final argon delivery system occurred in the fall of 2015, with cool-down of the acrylic vessel in the spring of 2016. The LAr fill began in the summer of 2016. After the neck seal failure incident on 17 August 2016, the AV was emptied of the LAr and refilled to a reduced level. Stable operations continue with a LAr mass of 3260~kg.

\section*{Acknowledgements}

This work is supported by the Natural Sciences and Engineering Research Council of Canada (NSERC), the Canadian Foundation for Innovation (CFI), the Ontario Ministry of Research and Innovation, and Alberta Advanced Education and Technology (ASRIP). We acknowledge support from the European Research Council Project ERC StG 279980, the UK Science \& Technology Facilities Council (STFC) grant ST/K002570/1, the Leverhulme Trust grant number ECF-20130496, the Rutherford Appleton Laboratory Particle Physics Division, and STFC and SEPNet PhD studentship support. We acknowledge support from DGAPA-UNAM through grant PAPIIT No. IA100316. We thank Compute Canada, Calcul Qu\'ebec, McGill University's centre for High Performance Computing and the Center for Advanced Computing (CAC) for computational support and data storage. 

We additionally thank SNOLAB and its staff for support through underground space, logistical, and technical services. SNOLAB operations are supported by the Canada Foundation for Innovation and the Province of Ontario Ministry of Research and Innovation, with underground access provided by Vale at the Creighton mine site. We thank Vale for their transportation of the acrylic vessel from surface to SNOLAB. On-site construction could not have been completed without the underground contractors, undergraduate research associates, and summer and co-op students who have made enormous contributions, and the management of Tony Flower. We thank the following people for their valuable inputs: David Bearse (Queen's), Neil Tennyson (Alfa Aesar), Dan Runciman (Johnston Industrial Plastics), Carlos Guerra (Spartech), Kalayil T. Varghese (RPT Asia) and Micha\l{} Tarka (Stony Brook). 

\bibliographystyle{elsart-num}
\bibliography{deap3600detectorMain}

\end{document}